\documentclass[showpacs,twocolumn,floatfix,amsmath,amssymb,prb]{revtex4}
\newif\ifpdf
\ifx\pdfoutput\undefined
   \pdffalse
\else
   \pdfoutput=1
   \pdftrue
\fi  

\ifpdf
  \usepackage[pdftex]{graphicx}
\else
  \usepackage[dvips]{graphicx}
\fi

\ifpdf
   \RequirePackage[pdftex, pdfpagemode=none, pdftoolbar=true,
                pdffitwindow=true, pdfcenterwindow=true]{hyperref}
      \usepackage{thumbpdf} 
\else
\fi

\usepackage{bm,times} 
\begin{document}
\title{Spin dynamics of the quasi two dimensional spin-$\frac{1}{2}$ 
quantum magnet Cs$_2$CuCl$_4$ } 
\author{ M.\ Y.\ Veillette, A.\ J.\ A.\ James and F.\ H.\ L.\ Essler }
\affiliation{Rudolf Peierls Centre for Theoretical Physics, University
of Oxford, 1, Keble Road, Oxford, OX1 3NP, United Kingdom\\}
\date{\today} 
\pacs{75.10.Jm, 75.25.+z,75.30.Ds.,75.40.Gb}

\begin{abstract}
We study dynamical properties of the anisotropic triangular quantum
antiferromagnet Cs$_2$CuCl$_4$. Inelastic neutron scattering
measurements have established that the dynamical spin correlations
cannot be understood within a linear spin wave analysis. We go beyond
linear spin wave theory by taking interactions between magnons into
account in a $1/S$ expansion. We determine the dynamical structure
factor and carry out extensive comparisons with experimental data. We
find that compared to linear spin wave theory a significant fraction
of the scattering intensity is shifted to higher energies and strong
scattering continua are present. However, the $1/S$ expansion fails to
account for the experimentally observed large quantum renormalization
of the exchange energies.
\end{abstract}
\maketitle
\section{Introduction}
\label{Introduction}
The quasi two dimensional spin-1/2 quantum magnet Cs$_{2}$CuCl$_{4}$
has attracted much theoretical and experimental interest in recent
years as a possible realization of a two dimensional quantum spin
liquid.~\cite{Coldea0,Coldea1,Coldea2,Coldea3,BETG,Bocquet,McKenzie1,Kim1,Zhang1,Zhang2,Wen1,Isakov}
This anisotropic triangular Heisenberg antiferromagnet is believed to
be a promising candidate due to its small spin, quasi two
dimensionality and geometrically frustrated spin
interactions. Although Cs$_{2}$CuCl$_{4}$ exhibits conventional
incommensurate long range magnetic order at low temperatures, neutron
scattering measurements have revealed unusual features in the spin
excitation spectrum. In particular, the dynamical correlations are
found to be dominated by an extended scattering continuum over a
relatively large window of energies. Several workers have interpreted
this observation as a signature of deconfined, fractionalized spin-1/2
(spinon) excitations, characteristic of a spin liquid phase. In this
line of approach, the observed broad scattering continuum is
interpreted in terms of a two-spinon scattering
continuum.~\cite{McKenzie1, Wen1, Kim1}

However, a strong scattering continuum does not entail an underlying
spin liquid phase. In fact, a conventional magnetically ordered phase
with strong magnon interactions can exhibit a broad continuum due to
multi magnon scattering processes. A previous examination of the
inelastic neutron scattering data on Cs$_2$CuCl$_4$ was performed in
the framework of linear spin wave (LSW) theory.~\cite{Coldea3} The
latter predicts sharp single particle excitations and weak two magnon
scattering continua, features which were argued to be in poor
agreement with the data. Given that the magnetic properties derive
from small $S=1/2$ Cu spins, one would {\it a priori} expect magnon
interactions to play an important role. In order to assess the
applicability of a spin wave based scenario to Cs$_2$CuCl$_4$ it is
therefore necessary to go beyond linear spin wave theory.

On a qualitative level the predictions of nonlinear spin wave theory
are readily anticipated. By Goldstone's theorem the breaking of a
continuous symmetry in a magnetically ordered state enforces the
presence of single particle excitations at low energies. As a result
of the aforementioned interactions, these magnons acquire a finite
life time, which in turn leads to a finite line width in the dynamical
structure factor. Furthermore, compared to linear spin wave theory,
spectral weight is transferred to higher energies via multi magnon
scattering processes. In the case of Cs$_{2}$CuCl$_{4}$ one may expect
the presence of a strong scattering continuum in the ordered phase
because (1) the low spin and the frustrated nature of the exchange
interactions lead to a small ordered moment and strong quantum
fluctuations around the ordered state; (2) the magnon interactions in
non-collinear spin structures like the ones found in
Cs$_{2}$CuCl$_{4}$ induce a coupling between transverse and
longitudinal spin fluctuations. This interaction provides an
additional mechanism for damping the spin waves and can enhance the
strength of the scattering continuum.

There is evidence of low-energy spin wave modes in the inelastic
neutron scattering data. Sharp peaks are also observed at high
energies near special wave vectors where a putative spin wave
dispersion is at a saddle-point. It is important to note that this
spin wave dispersion is dramatically ``renormalized'' compared to the
prediction of linear spin wave theory. \cite{Coldea2, Coldea3}

{\it A priori} it appears that nonlinear spin wave theory could have
the necessary ingredients to account for the spin correlations
observed in Cs$_2$CuCl$_4$. The issue then is whether it is possible
to achieve a {\em quantitative} description of the experiments in low
orders of perturbation theory in the spin wave interactions.

In the present work we go beyond linear spin wave theory and include,
within the framework of a $1/S$ expansion, the quantum fluctuations
around the classical ground state. We then apply the results to the
case $S=1/2$, in which the formal expansion parameter becomes of order
$1$ and is therefore not small. We are motivated by the observation
that spin wave theory gives a good description of physical properties
of the square-lattice spin-$\frac{1}{2}$ Heisenberg
Hamiltonian. \cite{Canali2, Hamer, Igarashi1} Indeed, higher order (in
a $1/S$ expansion) corrections to linear spin wave theory were shown
to be small in this case. Furthermore, taking these corrections into
account in the calculation of static and dynamical properties leads to
an improved agreement with the results of more sophisticated numerical
techniques. \cite{Singh2, Singh3, Sandvik} Although a corresponding
analysis is not available for the frustrated triangular
antiferromagnet, perturbative expansions in $1/S$ have shown the
renormalization due to quantum effects is relatively
small. \cite{Chubukov1,  Shiba2, Shiba1, Shiba3}
 
This paper is organized as follows. The spin Hamiltonian for
Cs$_2$CuCl$_4$ is introduced in Sec.~\ref{Spin Hamiltonian}. In
Sec.~\ref{Large S Expansion} we determine the magnon Green's function
in the framework of a large-S expansion. In Sec.~\ref{Dynamical
Correlation Function} we relate the experimentally measured dynamical
correlation functions to the magnon Green's function. The results of
our analysis and comparisons to the experimental data on
Cs$_2$CuCl$_4$ are presented in Sec.~\ref{Dynamical Properties of
Cs$_2$CuCl$_4$}. We conclude with a summary of our results in
Sec.~\ref{Conclusions}.

\section{Spin Model}
\label{Spin Hamiltonian}

\begin{figure}[ht]
\begin{center}
\includegraphics[width=8.6cm]{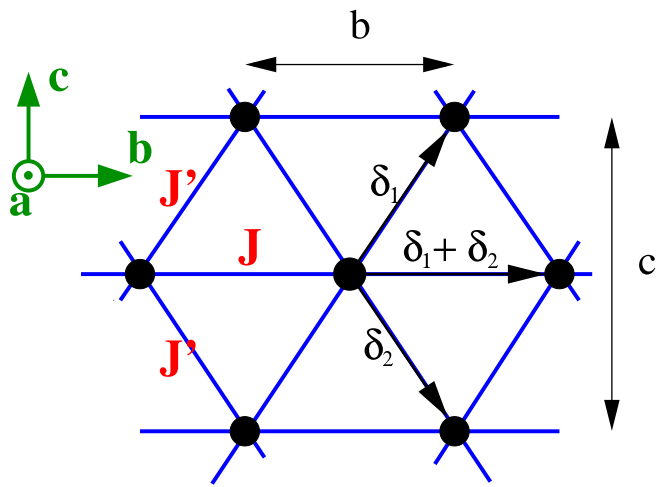}
\caption{(Color Online) The magnetic sites and exchange couplings
within a single layer of Cs$_{2}$CuCl$_{4}$. Layers are stacked along
the crystallographic $a$-direction with an interlayer spacing
${a}/{2}$ and a relative displacement in the $c$-direction.}
\label{Figure1}
\end{center}
\end{figure}

The full spin Hamiltonian of Cs$_{2}$CuCl$_{4}$ has been determined
previously from measurements in high magnetic fields (see
Ref.~\onlinecite{Coldea2} for details). For our purposes it suffices
to note that the magnetic Cu$^{2+}$ ions form a triangular lattice
with anisotropic exchange interactions. As shown in
Fig.~\ref{Figure1}, the main exchange interaction $J=0.374(5)$~meV is
along the crystallographic $b$ axis (``chain direction''). A weaker
spin exchange $J^{\prime}=0.128(5)$~meV occurs along the zig-zag
bonds. Finally, a Dzyaloshinskii-Moriya (DM) interaction
\cite{Dzyaloshinski58, Moriya60} $D=0.020(2)$~meV is present along the
zig-zag bonds.

Denoting the spin-$\frac{1}{2}$ operators at the sites ${{\bf R}}$ by
${\bf S}_{{\bf R}}$, the quasi two dimensional Hamiltonian takes the
form
\begin{widetext}
\begin{eqnarray}
\mathcal{H}=\sum_{{{\bf R}}} & J {\bf S}_{{\bf R}}\cdot {\bf S}_{{\bf
R}+\bf{\delta_{1}+{\delta_{2}}}} +J^{\prime} \left( {\bf S}_{{\bf
R}}\cdot {\bf S}_{{\bf R}+\bf{\delta_{1}}} + {\bf S}_{\bf R}\cdot {\bf
S}_{{\bf R}+\bf{\delta_{2}}} \right) - (-1)^n {\bf D } \cdot {\bf
S_{{\bf R}}}\times \left( {\bf S_{{\bf R}+\bf{\delta}_{1}}}+ {\bf
S_{{\bf R}+\bf{\delta}_{2}}} \right) .
\label{Hamiltonian0}
\end{eqnarray}
\end{widetext}
Here the vectors ${\bf \delta_{1}}$ and ${\bf \delta_{2}}$ connecting
neighboring sites are shown in Fig.~\ref{Figure1}. The vector ${\bf
D}=(D, 0, 0)$ is associated with the oriented bond between the two
coupled spins connected by $\bf{\delta}_1$ or $\bf{\delta}_2$ and $n$
is a layer index. The factor $(-1)^n$ indicates that the interaction
alternates between even and odd layers, which as a result can be
considered to be inverted versions of one another. A weak interlayer
interaction $J^{\prime \prime}$ is also present between neighboring
layers. However, as $J^{\prime\prime}$ is quite small we neglect it in
the following.

\begin{figure}[ht]
\includegraphics[width=8.6cm]{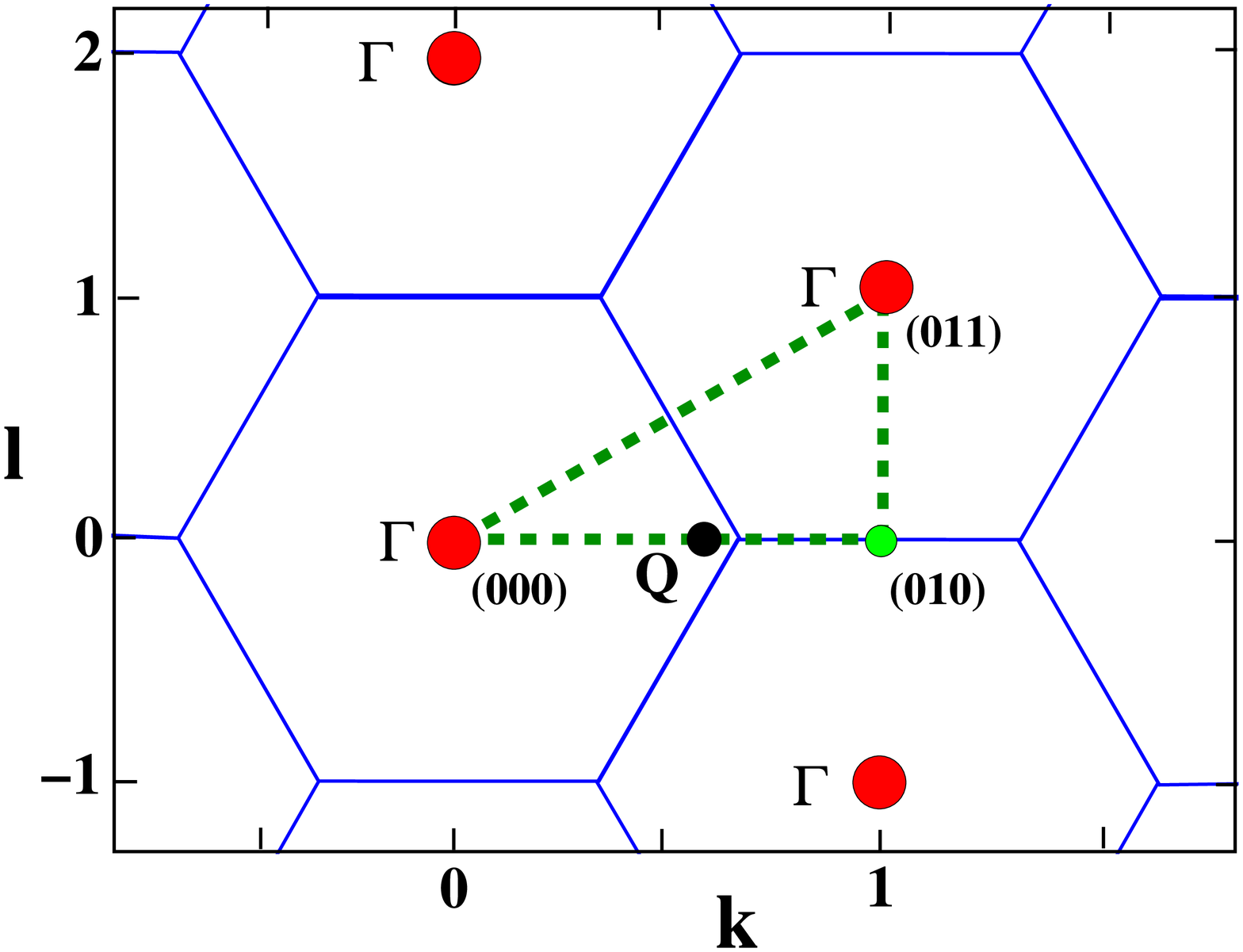} \\
\caption{(Color Online) The reciprocal space diagram of Cs$_2$CuCl$_4$
projected along the $(0, k, l)$ plane. The $\Gamma$ points refer to
the center of the Brillouin zone and ${\bf Q}$ is the ordering wave
vector. The path of the cut shown in Fig.~\ref{CutMain} is depicted as
a dashed line.}
\label{PATH}
\end{figure}

Following the conventions of Coldea {\em et al.} in
Ref.~\onlinecite{Coldea3}, we will discuss the dynamic response in
terms of the two dimensional Brillouin zone of the triangular lattice
even though the full crystal symmetry is orthorhombic. In our notation
wave vectors are expressed in terms of the reciprocal lattice vectors
as ${\bf k}=(h, k, l)$, which is a shorthand for $2\pi(h/a, k/b,
l/c)$.

The Fourier transforms of the exchange and DM interactions are
\begin{equation}
J_{\bf Q}= J \cos (2\pi k )+ 2 J^{\prime} \cos ( \pi k) \cos ( \pi l),
\end{equation}
and
\begin{equation}
D_{\bf Q}= - 2 i D \sin (\pi k) \cos( \pi l ).
\end{equation}
It is convenient for what follows to define a quantity
\begin{equation}
J^T_{\bf Q}= J_{\bf Q}-i D_{\bf Q}.
\end{equation}
Experimentally, spiral magnetic long range order is observed in
Cs$_2$CuCl$_4$ at temperatures below $T_N=0.62(1) K$. The ordered
structure is found to lie in the $bc$ plane by virtue of the small
easy-plane anisotropy generated by the DM interactions. The spin
structure is an incommensurate cycloid with an ordering wave vector
${\bf Q}=(0.0, 0.5+\epsilon, 0)$ where $\epsilon=0.030(2)$.

\section{Large S Expansion}
\label{Large S Expansion}

We now turn to a summary of our calculations. The procedure we follow
is standard. We first express the fluctuations around the
``classical'' ground state in terms of boson operators using the
Holstein-Primakoff transformation.~\cite{HolsteinPrimakoff,
Nagamiya, White01, Shiba2, Shiba1, Shiba3, Zhitomirsky02,
Zhitomirsky98} The term quadratic in the boson operators constitutes
the basis for linear spin wave theory, whereas higher order terms
represent spin wave interactions. The interaction vertices of $n$
bosons carry a factor $S^{2-n/2}$, where $S$ is the ``length'' of the
spin. In the second step we determine the renormalized magnon Green's
function by calculating the self-energy to leading order in
$1/S$. Finally, the experimentally observable dynamical correlation
functions are expressed in terms of the Green's function of the
Holstein-Primakoff bosons.

The classical ground state is determined by treating the spins as
classical vectors and then minimizing the energy. In this way one
obtains a cycloidal structure with a characteristic wave vector ${\bf
Q}$ that is fixed by the condition that it minimizes the exchange
energy per spin, i.e. $J^T_{\bf Q}={\rm min}_{\bf q}J^T_{\bf q}$. We
find ${\bf Q}=(0.0, 0.5+\epsilon_0, 0)$ with $\epsilon_0=0.054$. This
value differs significantly from the measured incommensuration but
quantum fluctuations lead to a reduction in $\epsilon_0$ and taking
them into account yields good agreement with
experiments.~\cite{Veillette, McKenzie1}

As we have already indicated in Eq.~(\ref{Hamiltonian0}), to a good
approximation the layers are decoupled. Hence we consider from now on
a set of independent 2-D layers, which are subdivided into two groups,
differing according to the direction of the DM vector. For the case
where the layer index $n$ is odd (even), the DM vector is taken to
point into (out of) the $bc$ plane.

In what follows we present the results for the even layers
only. However, it is easy to see that the spin structure factor is in
fact independent of the layer index and the overall result is a simple
summation over all layers.

It is convenient to define a local reference frame $(x, y, z)$ such
that the classical spin direction is aligned along the $z$ axis at
every site
\begin{eqnarray}
\begin{pmatrix}
S^a_{{\bf R}}\\ S^b_{{\bf R}}\\ S^c_{{\bf R}}
\end{pmatrix}
=
\begin{pmatrix}
1 & 0 &0 \\ 0 & \cos ({\bf Q} \cdot {{\bf R}}) & -\sin ({\bf Q} \cdot
{{\bf R}}) \\ 0 & \sin ({\bf Q} \cdot {{\bf R}}) & \cos ({\bf Q} \cdot
{{\bf R}})
\end{pmatrix}
\begin{pmatrix}
S^{x}_{{\bf R}}\\ S^{y}_{{\bf R}}\\ S^{z}_{{\bf R}}
\end{pmatrix}.
\label{rotation}
\end{eqnarray}
The Holstein-Primakoff transformation reads \cite{HolsteinPrimakoff}
\begin{eqnarray}
S^{+}_{{\bf R}}&=& S^{x}_{{\bf R}} + i S^{y}_{{\bf R}} = e^{i \theta}
\sqrt{\left( 2 S - \phi^{\dagger}_{{\bf R}}
\phi^{\phantom{\dagger}}_{{\bf R}} \right) }
\phi^{\phantom{\dagger}}_{{\bf R}}, \nonumber \\ S^{-}_{{\bf R}}&=&
S^{x}_{{\bf R}} - i S^{y}_{{\bf R}}= e^{-i \theta}
\phi^{\dagger}_{{\bf R}} \sqrt{ \left( 2 S - \phi^{\dagger}_{{\bf R}}
\phi^{\phantom{\dagger}}_{{\bf R}} \right)}, \nonumber \\ S^{z}_{\bf
R}&=& S - \phi^{\dagger}_{{\bf R}} \phi^{\phantom{\dagger}}_{{\bf R}},
\end{eqnarray}
where the boson creation and annihilation operators satisfy the
canonical commutation relation $\left [ \phi^{\phantom{\dagger}}_{\bf
R}, \phi^{\dagger}_{{{\bf R}}^\prime} \right]= \delta_{\bf {R,
R^{\prime}}}$. Here $\theta$ is an arbitrary angle which we set equal
to $\pi/2$ in order to make contact with the notation used in
Ref.~\onlinecite{Shiba1}. Introducing the Fourier transform
\begin{equation}
 \phi^{\dagger}_{{\bf k}}=\frac{1}{ \sqrt{N}} \sum_{{\bf R}}
 \phi^{\dagger}_{{\bf R}} e^{-i {\bf k} \cdot {{\bf R}}},
\end{equation}
on a lattice of $N$ sites, the Hamiltonian of Eq.~(\ref{Hamiltonian0})
takes the form
\begin{equation}
{\mathcal H}= {\mathcal H}_0+{\mathcal H}_2+{\mathcal H}_3 + {\mathcal
H}_4 +\cdots,
\label{expansionH}
\end{equation}
where ${\mathcal H}_{n}$ is proportional to $S^{2 -n/2}$ and consists
of normal ordered products of $n$ boson operators. There is no
${\mathcal H}_1$ term, because Eq.~(\ref{expansionH}) is an expansion
around a minimum of the classical energy. Linear spin wave theory
takes into account only the terms ${\mathcal H}_0$ and ${\mathcal
H}_2$. The higher order terms represent interactions between
magnons. The leading terms in the expansion are
\begin{widetext}
\begin{eqnarray}
{\mathcal H}_0 &=& N S^2 J^{T}_{\bf Q}, \\ {\mathcal H}_2 &=& N S
J^{T}_{\bf Q} + S \sum_{{\bf k}} A_{{\bf k}} \left( \phi^\dagger_{{\bf
k}} \phi^{\phantom{\dagger}}_{{\bf k}}+\phi^{\phantom{\dagger}}_{\bf
{ -k}} \phi^{{\dagger}}_{\bf {-k}} \right) -B_{{\bf k}} \left
( \phi^\dagger_{\bf {-k}} \phi^{{\dagger}}_{\bf
k}+\phi^{\phantom{\dagger}}_{\bf { -k}} \phi^{\phantom{\dagger}}_{\bf
k} \right) , \\ {\mathcal H}_3 &=& \frac{i}{2} \sqrt{\frac{S}{2 N}}
\sum_{{\bf 1}, {\bf 2}, {\bf 3}} \delta_{{\bf 1}+{\bf 2}+{\bf 3}}
\left( C_{\bf 1} + C_{\bf 2} \right) \left( \phi^{\dagger}_{-{\bf 3}}
\phi^{\phantom{\dagger}}_{{\bf 2}} \phi^{\phantom{\dagger}}_{{\bf 1}}
- \phi^{\dagger}_{\bf 1} \phi^{\dagger}_{{\bf 2}}
\phi^{\phantom{\dagger}}_{-{\bf 3}} \right), \\ {\mathcal H}_4 &=&
\frac{1}{4 N} \sum_{{\bf 1}, {\bf 2}, {\bf 3}, {\bf 4}} \delta_{{\bf
1}+{\bf 2}+{\bf 3}+{\bf 4}} \left\{ \frac{2}{3} \left( B_{{\bf 2}} +
B_{{\bf 3}} +B_{{\bf 4}} \right) \left( \phi^{\dagger}_{{\bf 1}}
\phi^{\phantom{\dagger}}_{-{\bf 2}} \phi^{\phantom{\dagger}}_{-{\bf
3}} \phi^{\phantom{\dagger}}_{-{\bf 4}}+ \phi^{\dagger}_{{\bf -4}}
\phi^{\dagger}_{{\bf -3}} \phi^{\dagger}_{{\bf -2}}
\phi^{\phantom{\dagger}}_{{\bf 1}} \right) \right. \nonumber \\ &+&
\left. \left[ \left( A_{{\bf 1}+{\bf 3}}+A_{{\bf 1}+{\bf 4}} +A_{{\bf
2}+{\bf 3}}+A_{{\bf 2}+{\bf 4}} \right) - \left( B_{{\bf 1}+{\bf 3}}
+B_{{\bf 1}+{\bf 4}}+B_{{\bf 2}+{\bf 3}}+B_{{\bf 2}+{\bf 4}} \right) -
\left( A_{{\bf 1}} +A_{{\bf 2}} +A_{{\bf 3}}+A_{{\bf 4}} \right)
\right] \phi^{\dagger}_{{\bf 1}} \phi^{\dagger}_{{\bf 2}}
\phi^{\phantom{\dagger}}_{-{\bf 3}} \phi^{\phantom{\dagger}}_{-{\bf
4}} \right\} .
\end{eqnarray}
\end{widetext}
Here the sum over ${\bf k}$ is performed in the first Brillouin zone
and the subscripts ${\bf 1} \ldots {\bf 4}$ denote ${\bf k}_{\bf 1}
 \ldots {\bf k}_{\bf 4}$. The quantities $A_{\bf k}, B_{\bf k}$ and
$C_{\bf k}$ are expressed as 
\begin{eqnarray}
 A_{{\bf k}}&=& \frac{1}{4}\left( 2J_{{\bf k}}+J^{T}_{\bf Q+ \bf k}+
 J^{T}_{\bf Q- \bf k} \right) - J^{T}_{\bf Q}, \nonumber \\ B_{\bf
 k}&=& \frac{1}{4}\left( 2 J_{{\bf k}}-J^{T}_{\bf Q + \bf k}-
 J^{T}_{\bf Q - \bf k}\right), \nonumber \\ C_{{\bf k}}&=& J^T_{\bf Q+
 \bf k} -J^T_{\bf Q- \bf k}.
\label{DefinitionsABC} 
\end{eqnarray}
The coefficients $A_{{\bf k}}$ and $B_{{\bf k}}$ are even functions of
${\bf k}$, whereas $C_{{\bf k}}$ is an odd function of ${\bf k}$. In
the absence of easy-plane anisotropies, i.e when $D$ vanishes and
inversion symmetry is present, we recover the results of
Ref.~\onlinecite{Shiba1}. [Note that our definitions in
Eqs.~(\ref{DefinitionsABC}) differ from those of
Ref.~\onlinecite{Shiba1} by a factor of four.] We emphasize that the
cubic interaction is generated as a result of the coupling between
transverse and longitudinal fluctuations and hence can only exist in
non-collinear spin structures. Furthermore, we note that the vertex
factor $C_{{\bf k}}\propto |{\bf {k}}|^3$ for small ${\bf k}$ owing to
the fact that $J^T_{\bf Q}$ is at a minimum by construction.

The quadratic Hamiltonian ${\mathcal H}_2$ is diagonalized by a
Bogoliubov transformation
\begin{eqnarray}
\phi^{\phantom{\dagger}}_{{\bf k}}&=& u_{{\bf k}}
\gamma^{\phantom{\dagger}}_{{\bf k}} + v_{{\bf k}}
\gamma^{\dagger}_{\bf {-k}}, \nonumber \\ \phi^{\dagger}_{-\bf k}&=&
v_{{\bf k}} \gamma^{\phantom{\dagger}}_{{\bf k}} + u_{{\bf k}}
\gamma^{\dagger}_{\bf {-k}},
\label{Bogoliubov}
\end{eqnarray}
where
\begin{eqnarray}
u^2_{{\bf k}}&=&1+v^2_{{\bf k}}= \frac{1}{2} \left( \frac{A_{{\bf
k}}}{\sqrt { A^2_{{\bf k}} -B^2_{{\bf k}}}} +1 \right), \nonumber \\
u_{{\bf k}} v_{{\bf k}} &=& \frac{1}{2} \frac{B_{{\bf k}}}{\sqrt {
A^2_{\bf k} -B^2_{{\bf k}}}}.
\label{Bogoliubov2}
\end{eqnarray}
The diagonal form of the quadratic Hamiltonian is
\begin{eqnarray}
{\mathcal H}_2 &=& N S J^{T}_{\bf Q} + \sum_{{\bf k}} \omega_{{\bf k}}
\left( \gamma^{\dagger}_{{\bf k}} \gamma^{\phantom{\dagger}}_{{\bf k}}
+ \frac{1}{2} \right) ,
\label{Bare}
\end{eqnarray}
where $\omega_{{\bf k}} = 2 S \sqrt{ A^2_{{\bf k}} -B^2_{{\bf k}}}$ is
the linear spin wave dispersion relation.~\cite{White01, Nagamiya} We
note that $\omega_{{\bf k}}$ is an even function of ${\bf k}$, despite
the absence of inversion symmetry in the Hamiltonian. In fact, the
symmetry of $\omega_{\bf k}$ is a consequence of time-reversal
symmetry, which implies the following relation between the elements of
the dynamical structure factor (Eq.~\ref{DSF}),~\cite{Lovesey}
\begin{equation}
S^{\mu\nu}_{{\bf k}, \omega}=S^{\nu\mu}_{\bf {-k}, \omega}.
\label{Tinv}
\end{equation}
The importance of quantum fluctuations can be gauged by determining
the average value of the local spin given by the standard formula
\begin{equation} 
\langle S^z_{\bf R} \rangle =S- \Delta S= S- \frac{1}{2 N} \sum_{\bf
k} u^2_{\bf k} +v^2_{\bf k}.
\label{DELTAS}
\end{equation}
The boson Green's function at zero temperature is expressed as
\begin{eqnarray}
G_{{\bf k}, \omega}= -i \int_{-\infty}^{\infty} d t e^{i \omega t}
 \bigg\langle T \left[
\begin{tabular}{l}
$\phi^{\phantom{\dagger}}_{{\bf k}} (t) $ \\ $\phi^{\dagger}_{\bf {
-k}} (t) $
\end{tabular} \right]
\left[ \phi^{\dagger}_{{\bf k}}(0) \phi^{\phantom{\dagger}}_{\bf {
-k}}(0) \right] \bigg\rangle , \nonumber\\
\end{eqnarray}
where $T$ denotes time ordering and $\langle ...\rangle$ represents a
ground state expectation value. The inverse of the unperturbed Green's
function is given by a $2 \times 2$ matrix,
\begin{equation}
G^{(0) -1}_{{\bf k}, \omega} = (-2 S A_{{\bf k}} + i \eta ) \sigma^0 +
2 S B_{{\bf k}} \sigma^x +\omega \sigma^z.
\label{Green}
\end{equation}
Here $\sigma^0$ and ${{\bm \sigma}}$ denote the identity and Pauli
matrices respectively and $\eta=0^+$.

The self-energy is defined by the Dyson equation,
\begin{equation}
G^{-1}_{{\bf k}, \omega} = G^{(0) -1}_{{\bf k},
\omega}-\Sigma^{\phantom{(0)}}_{{\bf k}, \omega},
\label{Green1}
\end{equation}
and can be parameterized as
\begin{equation}
\Sigma_{{\bf k}, \omega}= O_{{\bf k}, \omega} \sigma^0 + X_{{\bf k},
\omega} \sigma^x+Z_{{\bf k}, \omega} \sigma^z .
\end{equation}
The leading order (in $1/S$) contributions to the self-energy can be
divided into two parts
\begin{eqnarray}
\Sigma_{\bf k, \omega}= \Sigma^{(4)}_{{\bf k}} + \Sigma^{(3)}_{{\bf
k}, \omega}.
\end{eqnarray}

Here $\Sigma^{(4)}_{{\bf k}}$ denotes the vacuum polarization
contribution that arises in first order perturbation theory in
${\mathcal H}_4$. It is frequency independent and purely real. On the
other hand, $\Sigma^{(3)}_{{\bf k}, \omega}$ denotes the contribution
in second order perturbation theory of the three-magnon interaction
${\mathcal H}_3$. It incorporates the effects of magnon decay. Using
Eq.~(\ref{Green}), the $\Sigma^{(4)}_{{\bf k}}$ contribution to the
self-energy is found to be of the form
\begin{eqnarray}
O^{(4)}_{{\bf {\bf k}}}&=&A_{{\bf k}} +\frac{2 S}{N} \sum_{{\bf
k^{\prime}}} \frac{1}{\omega_{\bf k^{\prime}}} \bigg
[ \left(\frac{1}{2} B_{{\bf k}} +B_{\bf k^{\prime}} \right) B_{\bf
k^{\prime}} \nonumber \\ &+& \left( A_{{\bf k}-{\bf k^{\prime}}}
-B_{{\bf k}-{\bf k^{\prime}}} -A_{\bf k^{\prime}} -A_{\bf k} \right)
A_{\bf k^{\prime}} \bigg], \nonumber \\ X^{(4)}_{{\bf {\bf
k}}}&=&-B_{{\bf k}} +\frac{2 S}{N} \sum_{{\bf k^{\prime}}}
\frac{1}{\omega_{\bf k^{\prime}}} \bigg[ \left( B_{{\bf k}} +B_{\bf
k^{\prime}} \right) A_{\bf k^{\prime}} \nonumber \\ &+& \left(A_{{\bf
k}-{\bf k^{\prime}}} -B_{{\bf k}-{\bf k^{\prime}}} -A_{\bf k^{\prime}}
- \frac{1}{2} A_{{\bf k}} \right) B_{\bf k^{\prime}} \bigg], \nonumber
\\ Z^{(4)}_{{\bf k}}&=&0.
\label{Self2}
\end{eqnarray}
\begin{widetext}
The contribution $\Sigma^{(3)}$ is most easily evaluated in the
Bogoliubov basis ($\gamma$) and is equal to
\begin{eqnarray}
O^{(3)}_{{\bf k} , \omega}& =& \frac{-S}{16 N} \sum_{{\bf k^{\prime}}}
\left\{ \left[ \Phi^{(1)}( {\bf k^{\prime}}, {\bf k}-{\bf
k^{\prime}})\right]^2 +\left[ \Phi^{(2)}( {\bf k^{\prime}}, {\bf
k}-{\bf k^{\prime}}) \right]^2 \right\} \left( \frac{1}{ \omega_{\bf
k^{\prime}} + \omega_{{\bf k}-{\bf k^{\prime}}} - \omega - i \eta}+
\frac{1}{ \omega_{\bf k^{\prime}} + \omega_{{\bf k}-{\bf k^{\prime}}}
+ \omega - i \eta} \right), \nonumber \\ X^{(3)}_{{\bf k}, \omega} &=
& \frac{-S}{16 N} \sum_{{\bf k^{\prime}}} \left\{ \left[ \Phi^{(1)}
( {\bf k^{\prime}}, {\bf k}-{\bf k^{\prime}})\right]^2 -\left
[ \Phi^{(2)}( {\bf k^{\prime}}, {\bf k}-{\bf k^{\prime}}) \right]^2
\right\} \left( \frac{1}{ \omega_{\bf k^{\prime}} + \omega_{{\bf
k}-{\bf k^{\prime}}} - \omega - i \eta}+ \frac{1}{ \omega_{\bf
k^{\prime}} + \omega_{{\bf k}-{\bf k^{\prime}}} + \omega - i \eta}
\right), \nonumber \\ Z^{(3)}_{{\bf k}, \omega} &=& \frac{-S}{16 N}
\sum_{{\bf k^{\prime}}} \left\{ 2 \Phi^{(1)}( {\bf k^{\prime}}, {\bf
k}-{\bf k^{\prime}}) \Phi^{(2)}( {\bf k^{\prime}}, {\bf k}-{\bf
k^{\prime}})\right\} \left( \frac{1}{ \omega_{\bf k^{\prime}} +
\omega_{{\bf k}-{\bf k^{\prime}}} - \omega - i \eta}- \frac{1}
{ \omega_{\bf k^{\prime}} + \omega_{{\bf k}-{\bf k^{\prime}}} + \omega
- i \eta} \right),
\label{Self1}
\end{eqnarray}
where
\begin{eqnarray}
\Phi^{(1)}({\bf k^{\prime}}, {\bf k}-{\bf k^{\prime}}) &=& \left
( C_{\bf k^{\prime}} +C_{{\bf k}-{\bf k^{\prime}}} \right) \left
( u_{\bf k^{\prime}} +v_{\bf k^{\prime}} \right) \left( u_{{\bf
k}-{\bf k^{\prime}}} +v_{{\bf k}-{\bf k^{\prime}}} \right) -2 C_{{\bf
k}} \left( u_{{\bf k^{\prime}}} v_{{\bf k}-{\bf k^{\prime}}} +v_{{\bf
k^{\prime}}} u_{{\bf k}-{\bf k^{\prime}}} \right), \nonumber \\
\Phi^{(2)}({\bf k^{\prime}}, {\bf k}-{\bf k^{\prime}}) &=& C_{\bf
k^{\prime}} \left( u_{\bf k^{\prime}} +v_{\bf k^{\prime}} \right)
\left( u_{{\bf k}-{\bf k^{\prime}}} -v_{{\bf k}-{\bf k^{\prime}}}
\right)+C_{{\bf k}-{\bf k^{\prime}}} \left( u_{{\bf k}-{\bf
k^{\prime}}} +v_{{\bf k}-{\bf k^{\prime}}} \right) \left( u_{{\bf
k^{\prime}}} -v_{{\bf k^{\prime}}} \right).
\label{phi}
\end{eqnarray}
\end{widetext}
\section{Dynamical Correlation Function}
\label{Dynamical Correlation Function}

Inelastic neutron scattering experiments probe the dynamical structure
factor $S^{\mu\nu}_{{\bf k}, \omega}$. The latter is defined as the
Fourier transform of the dynamical spin-spin correlation function
\begin{equation}
S^{\mu \nu}_{{\bf k}, \omega}=\int_{-\infty}^{\infty} \frac{d t}{2 \pi
} e^{- i \omega t } \langle S^{\mu}_{-{\bf k}} (0) S^{\nu}_{{\bf
k}}(t) \rangle .
\label{DSF}
\end{equation}
Here $\mu, \nu=(a, b, c)$ label the various crystallographic axes and
the Fourier-transformed spin operators are defined by $S^{\mu}_{\bf
k}=\frac{1}{\sqrt{N}} \sum_{{\bf R}} S^{\mu}_{{\bf R}} e^{-i {\bf k}
\cdot {{\bf R}}}$.

It is convenient to introduce time-ordered spin-spin correlation
functions in the rotated coordinate system
\begin{equation}
F^{\alpha \beta}_{{\bf k}, \omega}= -i \int_{-\infty}^{\infty} d t
e^{- i \omega t } \langle T S^{\alpha}_{-{\bf k}} (0) S^{\beta}_{{\bf
k}}(t) \rangle ,
\end{equation}
where $\alpha, \beta=(x, y, z)$ are the rotated coordinate axes
(Eq.~\ref{rotation}). The dynamical structure factor is related to the
imaginary part of the time ordered correlation function in the
following way
\begin{eqnarray}
S^{aa}_{{\bf k}, \omega}&=& -\frac{1}{\pi} {\rm Im} F^{x x}_{{\bf k},
\omega}, \label{Saa} \\ S^{bb}_{{\bf k}, \omega}&=& S^{cc}_{{\bf k},
\omega}= -\frac{1}{\pi} {\rm Im} \left[ \Theta^{+}_{{\bf k}+{\bf Q},
\omega} + \Theta^{-}_{{\bf k}-{\bf Q}, \omega} \right], \label{Sbb}\\
S^{bc}_{{\bf k}, \omega}&=& -S^{cb}_{{\bf k}, \omega}= -\frac{i}{\pi}
{\rm Im} \left [ \Theta^{+}_{ {\bf k}+{\bf Q}, \omega} -
\Theta^{-}_{{\bf k}-{\bf Q}, \omega } \label{Scb} \right],
\end{eqnarray}
where
\begin{equation}
\Theta^{\pm}_{{\bf k}, \omega}= \frac{1}{4} \left[ F^{z z}_{{\bf k},
\omega} +F^{y y}_{{\bf k}, \omega} \pm i \left( F^{z y}_{{\bf k},
\omega} - F^{y z}_{{\bf k}, \omega} \right) \right].
\end{equation}

To proceed further, we expand the dynamical correlation functions in
inverse powers of $S$ to order ${\cal O}(S^0)$. The corresponding
results have been derived previously by Ohyama and
Shiba.~\cite{Shiba1} Here we merely quote their results for the sake
of completeness. The transverse correlations are
\begin{eqnarray}
F^{x x}_{{\bf k}, \omega}&=& \frac{S}{2} c^2_{x} {\rm Tr} \left[\left(
\sigma^0- \sigma^x \right) G_{{\bf k}, \omega} \right], \nonumber \\
F^{y y}_{{\bf k}, \omega}&=& \frac{S}{2} c^2_{y} {\rm Tr} \left[\left(
\sigma^0+ \sigma^x \right) G_{{\bf k}, \omega} \right],
\end{eqnarray}
where the Green's function is given by Eq.~(\ref{Green1}) and where
\begin{eqnarray}
c_{x}&=& 1- \frac{1}{ 4 S N} \sum_{{\bf k}} \left( 2 v^2_{{\bf k}}-
u_{{\bf k}} v_{{\bf k}} \right), \nonumber \\ c_{y}&=& 1- \frac{1}{ 4
S N} \sum_{{\bf k}} \left( 2 v^2_{{\bf k}}+ u_{{\bf k}} v_{{\bf k}}
\right).
\label{cxy}
\end{eqnarray}
We note that when squaring~(\ref{cxy}) only terms to order ${\cal
O}(S^{-1})$ must be retained. The mixing of transverse and
longitudinal fluctuations manifests itself in
\begin{eqnarray}
i \left( F^{y z}_{{\bf k}, \omega} -F^{z y}_{ {\bf k}, \omega} \right)
=&& c_{y} \left\{ P^{(1)}_{{\bf k}, \omega} {\rm Tr} \left[\left (
\sigma^0 + \sigma^x \right) G_{{\bf k}, \omega} \right]
\right. \nonumber \\ && \qquad + \left. P^{(2)}_{{\bf k}, \omega} {\rm
Tr} \left[\sigma^z G_{{\bf k}, \omega} \right] \right\}.
\end{eqnarray}
Here the functions $P^{(1, 2)}_{{\bf k}, \omega}$ are defined as
\begin{widetext}
\begin{eqnarray}
P^{(1)}_{{\bf k}, \omega}= &\frac{S}{4 N}& \sum_{{\bf k^{\prime}}}
\Phi^{(1)} \left( {\bf k^{\prime}}, {\bf k}-{\bf k^{\prime}} \right)
\left( u_{{\bf k^{\prime}}} v_{{\bf k} -{\bf k^{\prime}}} + v_{{\bf
k^{\prime}}} u_{{\bf k}-{\bf k^{\prime}}} \right)
\left( \frac{1}{ \omega_{\bf k^{\prime}} + \omega_{{\bf k}-{\bf
k^{\prime}}} - \omega - i \eta}+ \frac{1}{ \omega_{\bf k^{\prime}} +
\omega_{{\bf k}-{\bf k^{\prime}}} + \omega - i \eta} \right),
\nonumber \\ P^{(2)}_{{\bf k}, \omega} = &\frac{S}{4 N}& \sum_{{\bf
k^{\prime}}} \Phi^{(2)} \left( {\bf k^{\prime}}, {\bf k}-{\bf
k^{\prime}} \right) \left( u_{{\bf k^{\prime}}} v_{{\bf k} -{\bf
k^{\prime}}} + v_{{\bf k^{\prime}}} u_{{\bf k}-{\bf k^{\prime}}}
\right)
\left( \frac{1}{ \omega_{\bf k^{\prime}} + \omega_{{\bf k}-{\bf
k^{\prime}}} - \omega - i \eta}- \frac{1}{ \omega_{\bf k^{\prime}} +
\omega_{{\bf k}-{\bf k^{\prime}}} + \omega - i \eta} \right).
\label{P1P2}
\end{eqnarray}
Finally the longitudinal correlations are decomposed in inverse powers
of $S$ as $F^{ z z}_{{\bf k}, \omega} = F^{(0) z z}_{{\bf k},
\omega}+F^{(1) z z}_{{\bf k}, \omega}$, where
\begin{eqnarray}
&&F^{(0) z z}_{{\bf k}, \omega}=-\frac{1}{2 N} \sum_{{\bf k^{\prime}}}
\left( u_{{\bf k^{\prime}}} v_{{\bf k} -{\bf k^{\prime}}} + v_{{\bf
k^{\prime}}} u_{{\bf k}-{\bf k^{\prime}}} \right)^2
\left( \frac{1}{ \omega_{\bf k^{\prime}} + \omega_{{\bf k}-{\bf
k^{\prime}}} - \omega - i \eta}+ \frac{1}{ \omega_{\bf k^{\prime}} +
\omega_{{\bf k}-{\bf k^{\prime}}} + \omega - i \eta} \right),
\label{Twomagnonzeroorder}\\
&&F^{(1) z z }_{{\bf k}, \omega}= \frac{1}{2 S} \left\{ \left(
P^{(1)}_{{\bf k}, \omega} \right)^2 {\rm Tr} \left[\left( \sigma^0 +
\sigma^x \right) G_{{\bf k}, \omega} \right]+ \left( P^{(2)}_{{\bf k},
\omega} \right)^2 {\rm Tr} \left[\left( \sigma^0 - \sigma^x \right)
G_{{\bf k}, \omega} \right]
+2 P^{(1)}_{{\bf k}, \omega} P^{(2)}_{{\bf k}, \omega} {\rm Tr}
\left[\sigma^z G_{{\bf k}, \omega} \right] \right\}.
\label{Longitudinal}
\end{eqnarray}
\end{widetext}
We note that the $F^{(0) z z}$ term does not require the knowledge of
the bosonic self-energy and is basically a free boson result. For this
reason, it is often included in linear spin wave calculation as a
source of two magnon scattering, even though it is formally a higher
order contribution in $1/S$. In what follows, we abide by this (in
some sense inconsistent) convention and consider the contribution of
Eq.~\ref{Twomagnonzeroorder} as part of linear spin-wave theory. As a
consequence we then retain the $F^{(1) z z}$ contribution to the
dynamical structure factor, although of higher order in $1/S$
(i.e. ${\cal O}(S^{-1})$) than the other terms we take into account.

The (unpolarized) inelastic neutron scattering cross section is given
by
\begin{eqnarray} 
\frac{d^2 \sigma}{d \omega d \Omega} &=& |f_{{\bf k}}|^2 \sum_{ \mu
\nu} \left ( \delta_{\mu \nu}- \hat{{\bf k}}_{\mu} \hat{{\bf k}}_{\nu}
\right) S^{\mu \nu}_{{\bf k}, \omega} , \nonumber \\ &=&|f_{{\bf
k}}|^2 \left[ \bigl( 1- \hat{{\bf k}}^2_{a} \bigr) S^{aa}_{{\bf k},
\omega} + \bigl(1+ \hat{{\bf k}}^2_{a} \bigr) S^{bb}_{{\bf k}, \omega}
\right],
\label{crosssection}
\end{eqnarray}
where $\hat{{\bf k}}_{\mu}$ is the $\mu$-component of the unit vector
in ${\bf k}$ direction. The magnetic form factor $f_{{\bf k}}$ is
determined by the magnetic ions. For Cu$^{2+}$, the isotropic form
factor has a relatively weak wave vector dependence within the first
Brillouin zone and will be neglected from now on.~\cite{Wilson}

It is well known that the $1/S$ expansion preserves many physical
properties ``order by order'' in $1/S$. For instance, it follows from
Eqs.~(\ref{Self2}, \ref{Self1}) that the Goldstone modes persist
beyond linear spin wave theory as one expects on physical grounds. A
careful examination also shows that to order ${\cal O}(S^0)$ the
spectral functions are positive and that the relation~(\ref{Tinv})
holds. However, due to a lack of self-consistency the $1/S$ expansion
leads to an (unphysical) unequal treatment of the one-magnon and two
magnon scattering contributions to dynamical correlation
functions.~\cite{Shiba3} It is worthwhile to discuss this issue in
more detail. The leading order contribution to the dynamical structure
factor is due to coherent single magnon excitations and is of the form
$\delta \left ( \omega -\omega_{{\bf k}} \right)$. The two magnon
contribution due to longitudinal fluctuations
(Eq.~\ref{Twomagnonzeroorder}) gives rise to a scattering continuum of
the form $\sum_{\bf k^{\prime}} I \left( {\bf k}, {\bf k^{\prime}}
\right) \delta \left( \omega -\omega_{\bf k^{\prime}}-\omega_{{\bf
k}-{{\bf k}^{\prime}}} \right)$ with some function $I \left( {\bf k},
{\bf k^{\prime}} \right)$. The extent of the two magnon contribution
in ${\bf k}-\omega$ space is determined by the lower and upper bounds
of the function $\omega_{\bf k^{\prime}}+\omega_{{\bf k}-{{\bf
k}^{\prime}}}$ for a given ${\bf k}$.

On general grounds, we expect the lower bound of the two magnon
scattering continuum to be equal to or smaller than the ``true''
magnon dispersion $\bar{\omega}_{{\bf k}}$. In fact, the existence of
a zero-momentum Goldstone mode guarantees that there exists a two
magnon contribution at frequencies $\bar{\omega}_{{\bf k}}+
\bar{\omega}_{\bf 0} =\bar{\omega}_{{\bf k}}$.

It is easy to see that this property does {\sl not} hold order by
order in a $1/S$ expansion. Indeed, the first order contribution in
$1/S$ shifts the pole of the Green's function and leads to a
renormalization of the magnon dispersion. The renormalized dispersion
$\tilde{\omega}_{{\bf k}}$ can be determined from the Dyson equation
\begin{equation}
G^{-1}_{{\bf k}, \tilde{\omega}_{{\bf k}}} = G^{(0) -1}_{{\bf k},
\tilde{\omega}_{{\bf k}}}-\Sigma^{\phantom{-1}}_{{\bf k},
\tilde{\omega}_{{\bf k}}}=0.
\label{Dyson}
\end{equation}
However, to order ${\cal O}(S^0)$ the threshold of the two magnon
contribution is still determined by the bare dispersion relation
$\omega_{{\bf k}}$. This results in an unphysical behavior, where the
two magnon scattering continuum is separated from the single magnon
dispersion by a gap. In order to avoid this problem, we impose the
following self-consistency condition: {\sl the linear spin wave
dispersion $\omega_{{\bf k}}$ used in Eqs.~(\ref{Self1}, \ref{P1P2},
\ref{Twomagnonzeroorder}) is to be replaced by the renormalized
dispersion $\tilde{\omega}_{{\bf k}}$.}
\begin{figure*}[ht]
\includegraphics[width=18cm]{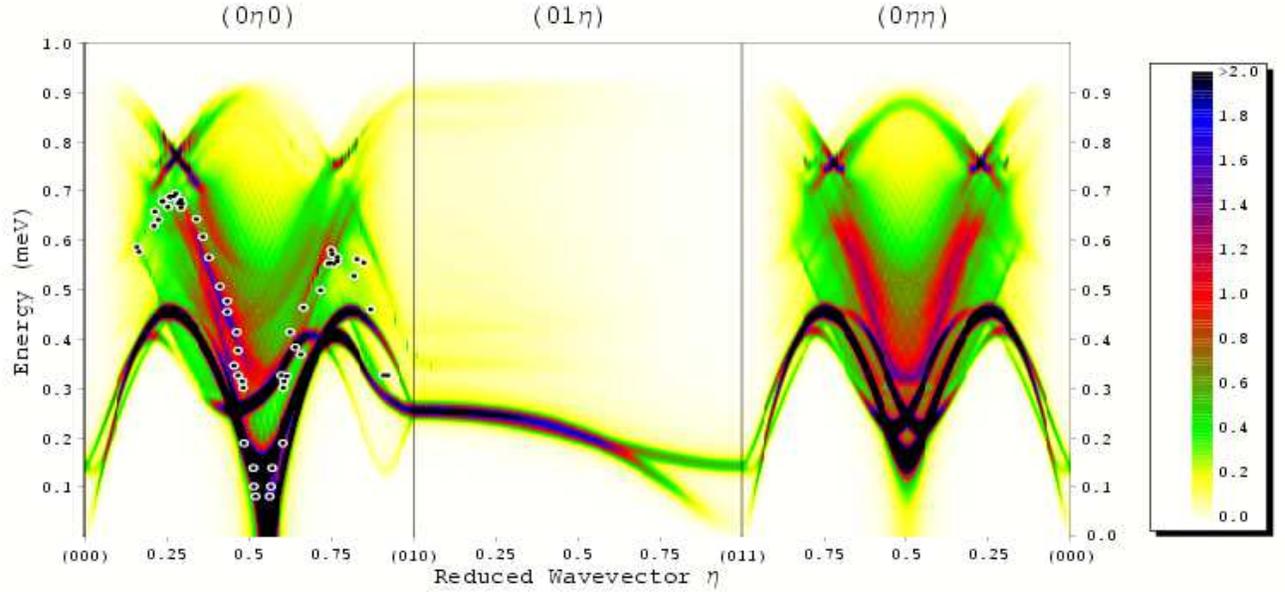}
\caption{(Color Online) Density plot of the scattering cross-section
(Eq.~\ref{crosssection}) as a function of energy and wave vector. The
light (dark) regions represent regions of small (large) scattering
intensity. The results have been convolved with the experimental
energy resolution of the detectors (The full width at half maximum is
$\Delta E=0.016$ meV). The magnetic form factor of copper in
Eq.~(\ref{crosssection}) shows very weak wave vector dependence in the
regime of interest and therefore was taken to be unity.~\cite{Wilson}
The filled circles along the (010) direction are the experimental
position of the most intense peaks in the line shapes taken in the
spiral phase ($T<0.1$K).~\cite{Coldea3}}
\label{CutMain}
\end{figure*}
\begin{figure}[h2t]
\includegraphics[width=8.6cm]{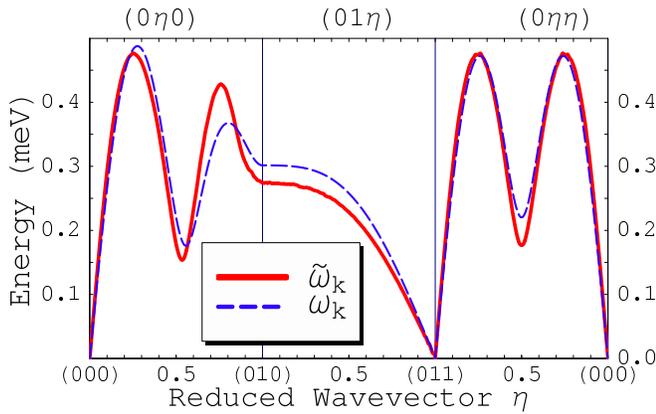}
\caption{(Color Online) The renormalization of the spin wave
spectrum. The solid and dashed lines are results of the $1/S$
expansion $\tilde{\omega}_{{\bf k}}$ (Eq.~\ref{Dyson}) and the linear
spin wave dispersion $\omega_{{\bf k}}$ (Eq.~\ref{Bare}). The cut in
the paramagnetic Brillouin zone runs from $(000)$ to $(010)$ to
$(011)$ and back to the center of the Brillouin zone $(000)$.}
\label{Spectrum}
\end{figure}
\begin{figure}[ht]
\begin{picture}(8, 5)(0.25, 0)
\put(0, 0){\includegraphics[width=8.6cm]{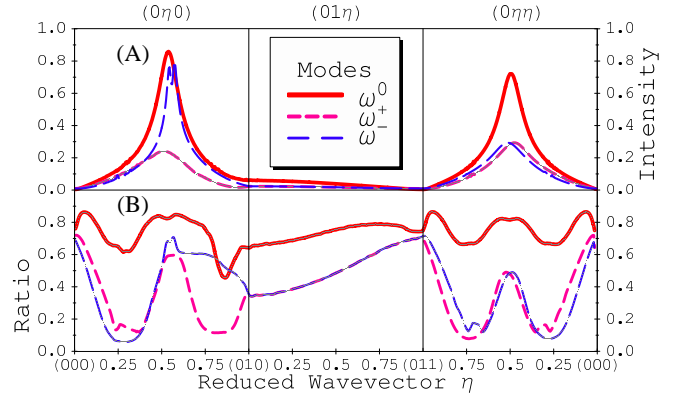}} \put(1.6,
4.5){\makebox(0, 0){(A)}} \put(1.6, 2.5){\makebox(0, 0){(B)}}
\end{picture}
\caption{(Color Online) Upper panel (A): The integrated spectral
weights for the three polarizations (Eq.~\ref{Intensity}) as functions
of momentum transfer. The principal $\omega^{0}$ and secondary
$\omega^{+}$, $\omega^{-}$ spin wave modes are defined in the
text. Lower panel (B): The ratios of the spectral weights of the
single-particle excitations of each polarization to their respective
integrated intensities (Eq.~\ref{Ratio}).}
\label{CutMainCOMBO}
\end{figure}
\section{Dynamical Properties of $\mbox{Cs}_2\mbox{CuCl}_4$}
\label{Dynamical Properties of Cs$_2$CuCl$_4$}

So far our discussion of the $1/S$ expansion has been fairly
general. In order to make contact with the experiments on
Cs$_2$CuCl$_4$ we now set the exchange constants to their appropriate
values \cite{Coldea2, Coldea3} and fix $S=1/2$. We then evaluate the
dynamical structure factor at a given wave vector numerically. Complex
integrals such as Eqs.~(\ref{Self1}, \ref{P1P2}) are evaluated by
summing the imaginary part of the integrands over a frequency grid of
1200 points and of 1000 $\times$ 1000 points in wave vector space. The
real parts are then determined from the Kramers-Kronig relations. The
aforementioned self-consistency condition is implemented by
calculating the full Green's function iteratively on a 100 $\times$
100 grid in the Brillouin zone. We observe satisfactory numerical
convergence after about three iterations.

We first turn to the magnon dispersion. The linear spin wave result
$\omega_{{\bf k}}$ vanishes at the center of the paramagnetic
Brillouin zone. The corresponding Goldstone mode is associated with
small fluctuations of the ordered moment within the cycloidal
plane. In helimagnets, the spectrum often exhibits a second Goldstone
mode at the ordering wave vector. This gapless mode is due to
fluctuations of the plane of the cycloid. In the case at hand, the
easy-plane anisotropy generated by the DM term forces the cycloidal
structure to lie in the $bc$ plane and creates an excitation gap at
the ordering wave vector ${\bf Q}$.

The renormalization of the magnon dispersion within the framework of
the $1/S$ expansion is obtained from the poles of the Green's function
(Eq.~\ref{Dyson}). In Fig.~\ref{Spectrum} we compare the results of
the $1/S$ expansion with the linear spin wave theory. It is customary
to quantify the effects of the ``quantum'' renormalization of the
magnon dispersion by parametrizing the latter in terms of
``effective'' exchange constants $ \tilde{J}, \tilde{J}^{\prime},
\tilde{D}$ and comparing them with the ``bare'' parameters $J$,
$J^\prime$ and $D$.

Experimentally, the quantum renormalization is found to be rather
large, namely $\frac{\tilde{J}}{J}=1.63(5)$ and
$\frac{\tilde{J}^{\prime}}{J^{\prime}}=0.84(9)$. The renormalization
of $D$ was not established. The $1/S$ expansion yields the
significantly smaller renormalizations $\frac{\tilde{J}}{J}=1.131$,
$\frac{\tilde{J}^{\prime}}{J^{\prime}}=0.648$ and
$\frac{\tilde{D}}{D}=0.72$. The difference between the theoretical and
experimental values indicates that the leading order in a $1/S$
expansion underestimates fluctuation effects. On the other hand one
should note that the $1/S$ expansion gives a result of $0.031$ for the
incommensuration, which is very close to the experimentally observed
value of $\epsilon=0.030(2)$.

Before turning to a comparison of our results for the dynamical
structure factor with the experimental results, we briefly review some
facts about excitations in helimagnets. Generally it is useful to
distinguish between three spin wave modes. In the case at hand, the
``principal'' mode $\omega^0_{{\bf k}}=\tilde{\omega}_{{\bf k}}$ is
polarized along the $a$ axis and is probed by the $S^{aa}_{{\bf k},
\omega}$ component of the dynamical structure factor
(Eq.~\ref{Saa}). The two ``secondary'' spin wave modes 
$\omega^{\pm}_{{\bf k}}=\tilde{\omega}_{\bf k \pm \bf Q}$  are images
of the main mode but their momenta are shifted by $\pm {\bf Q}$. They
are polarized in the $bc$ plane (Eqs.~\ref{Sbb}, \ref{Scb}). In linear
spin wave theory, the three spin wave modes give rise to sharp
$\delta$ functions and carry a large part of the spectral weight.

In addition to the single magnon modes there are multi magnon
scattering continua. Whenever the magnon dispersion lies within a
scattering continuum, the single magnon excitation gets broadened and
acquires a finite line width. On the other hand, when the magnon
dispersion lies at the threshold of a scattering continuum, there is
no significant decay and the single magnon mode remains sharp.

The unpolarized dynamical structure factor (where the various
polarizations are added according to Eq.~\ref{crosssection}) is shown
in Fig.~\ref{CutMain} as a function of energy and momentum for a
particular ``cut'' of momentum transfers. The cut along the $b$
direction, i.e. from $(000)$ to $(010)$, shows large modulations of
the dispersion relation due to the strong intra-chain
correlations. Near the ordering wave vector ${\bf Q}$ the scattering
intensity increases sharply. For momentum transfers perpendicular to
the chains, (i.e. along the $(01\eta)$ direction), the single particle
modes are seen to be resolution limited. The two in-plane modes become
degenerate and their dispersions are nearly featureless, whereas the
out-of-plane fluctuations dip to zero energy at $(011)$, in accordance
with Goldstone's theorem. Along the $(0\eta \eta)$ direction the
spectrum is symmetric across the Brillouin zone boundary. Additional
structures due to two magnon scattering are clearly visible at higher
energies along the $(0\eta0)$ and $(0 \eta \eta)$ directions.

In order to illustrate how the spectral weights associated with the
single-particle excitations are affected by the magnon interactions,
we have estimated their contributions for each polarization to the
integrated spectral weights. The total integrated intensity of each
polarization is given by ``equal-time'' correlation functions,
\begin{eqnarray}
 I^0_{{\bf k}} &=& -\frac{1}{\pi} \int d \omega \thinspace
F^{xx}_{{\bf k}, \omega}, \nonumber \\ I^\pm_{{\bf k}} &=&
-\frac{1}{\pi} \int d \omega \thinspace \Theta^{\pm}_{{\bf k \pm \bf
Q}, \omega}.
\label{Intensity}
\end{eqnarray} 
The one-magnon contribution to the integrated intensity of each
polarization is then determined by integrating the respective
correlation function in the vicinity of the single particle
dispersions. In practice we find that integrating the peaks assuming a
Lorentzian form is a poor prescription for strongly damped
peaks. Instead we numerically integrate the intensity over an energy
window of three times the width at half maximum
\begin{eqnarray}
 R^0_{{\bf k}} &=& \frac{1}{ I^0_{\bf k}} \int^{\omega^{0}_{\bf k}+1.5
\Delta \omega^0_{\bf k}}_{\omega^{0}_{\bf k}-1.5 \Delta
\omega^{0}_{\bf k}} d \omega \thinspace \frac{-F^{xx}_{{\bf
k},\omega}}{\pi}, \nonumber \\ R^{\pm}_{{\bf k}} &=& \frac{1}
{ I^{\pm}_{\bf k}} \int^{\omega^{\pm}_{{\bf k} \pm {\bf Q}}+1.5 \Delta
\omega^{\pm}_{{\bf k}\pm {\bf Q}}}_{\omega^{\pm}_{{\bf k}\pm {\bf Q}}
-1.5 \Delta \omega^{\pm}_{{\bf k} \pm {\bf Q}}} d \omega \thinspace
\frac{-\Theta^{\pm}_{{{\bf k}\pm {\bf Q}},\omega}}{\pi}.
\label{Ratio}
\end{eqnarray} 
The results are shown in Fig.~\ref{CutMainCOMBO}. We see that the
integrated spectral weight is concentrated in the vicinities of the
ordering wave vector ${\bf Q}$ and $(0 \frac{1}{2} \frac{1}{2})$ and
is largely suppressed near the $\Gamma$ point. The weights associated
with single magnon excitations are strongly suppressed for the
secondary modes. This is a consequence of the non-collinearity of the
magnetic order. The in-plane modes are significantly damped as a
result of the coupling between longitudinal and transverse
fluctuations. Such a coupling is not present for the out-of-plane mode
and therefore the principal mode carries generally more spectral
weight. Nevertheless the fraction of spectral weight associated with
single-particle excitations decreases significantly whenever the
renormalized spin wave dispersion is pushed upwards in energy for a
given momentum. For instance, near wave vector $(0, 0.8, 0)$ the
principal spin wave mode lies within the two magnon continuum and as a
result less than 50 \% of the spectral weight is attributed to the
one-magnon excitation.

\begin{figure}[ht]
\begin{center}
\includegraphics[width=8.6cm]{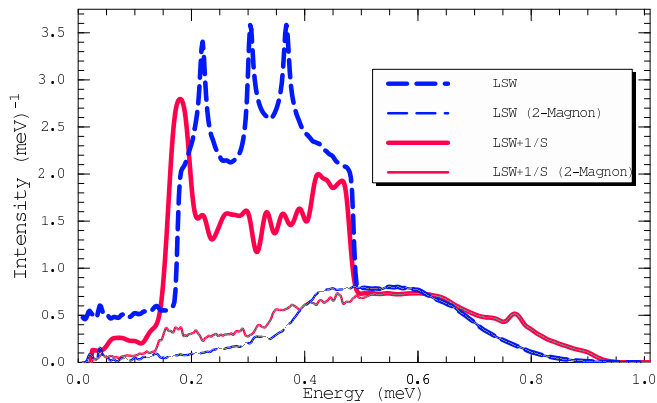}
\caption{(Color Online) The scattering intensity as a function of
energy (Eq.~\ref{IOMEGA}) for LSW theory (dashed line) and LSW+1/S
expansion (solid line). The contributions of the scattering continua
is shown using thin lines.}
\end{center}
\label{AVEI}
\end{figure}

The scattering intensity can also be studied by performing a wave
vector average,
\begin{equation}
I_{T} (\omega)= \frac{1}{N} \sum_{\bf k}\sum_{\mu} S^{\mu \mu}_{{\bf
k}, \omega}.
\label{IOMEGA}
\end{equation}
By the frequency sum rule, the scattering intensity (Eq.~\ref{IOMEGA})
integrated over all energies (including the elastic Bragg peaks at
$\omega=0$) has to equal $S(S+1)$. However, this sum rule does not
hold ``order by order'' in perturbation theory. For instance, the
total intensity within linear spin wave theory exceeds the sum rule by
$ \Delta S (1 +2 \Delta S) $. Bearing this caveat in mind, the sum
rule is a useful tool for comparing the one and two magnon
contributions as well as analyzing the shift in spectral weight. In
Fig.~\ref{AVEI}, we plot the scattering intensities as functions of
energy within linear spin wave theory and the $1/S$ expansion. In
linear spin wave theory the integrated intensity exhibits cusps, which
are associated with van-Hove singularities in the single particle
density of states. In the $1/S$ expansion such sharp features are
absent. Above approximately $0.5$~meV the one magnon contribution
vanishes and the scattering intensity is entirely due to multi magnon
states.

To quantify the shift of the spectral weight we calculate the first
moment of the normalized scattering intensity $\langle \omega
\rangle$. We find that the linear spin wave theory value $\langle
\omega \rangle = 0.35$~meV is renormalized upwards to $\langle
\omega \rangle = 0.40$~meV in the $1/S$ expansion. This
observation is in line with the expectation that the higher orders of
the $1/S$ expansion induce a transfer of spectral weight to higher
energies via multi magnon scattering processes. In fact, as shown in
Fig.~\ref{AVEI} the two magnon contribution to the overall intensity
is $29\%$ in linear spin wave theory but $46 \%$ in the $1/S$
expansion.

\begin{figure}[ht]
\begin{center}
\includegraphics[width=8.6cm]{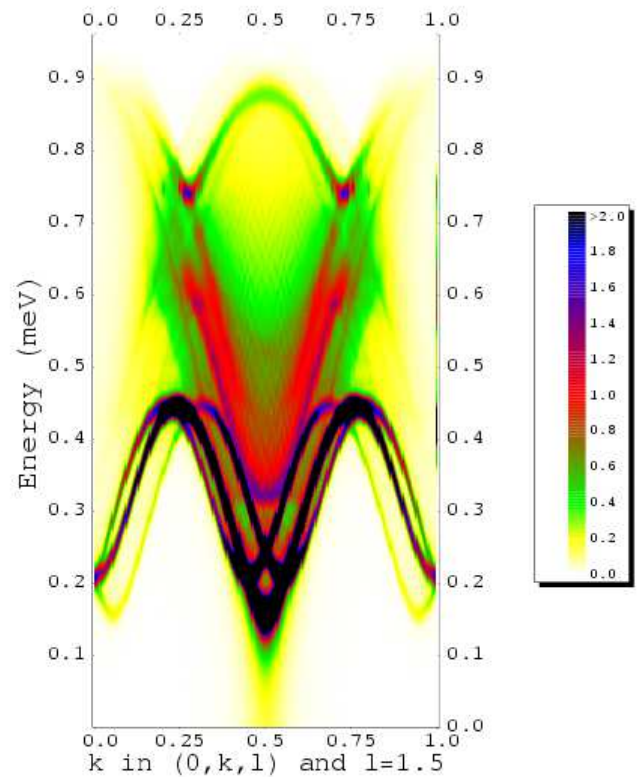}
\caption{(Color Online) Density plot of the scattering cross-section
as a function of energy and wave vector along the cut
$(0k\frac{3}{2})$. The light (dark) regions represent regions of small
(large) scattering intensity.}
\end{center}
\label{CutB}
\end{figure}

\begin{table}
\begin{center}
\begin{ruledtabular}
\begin{tabular}{cccc}
\# &$k$(rlu)&$l$(rlu)&$\hat{\bf k}_a$\\
\hline
A & $-0.389 +0.189E-0.016E^2$ & $0 $ & 0.05 \\ B & $-0.30
+0.189E-0.015E^2$ & $0 $ & 0.02 \\ C & $ 0.21 +0.297E -0.026E^2$ & $0
$ & 0.25 \\ D & $2.11 +0.29E -0.025E^2$ & $0$ & 0.95 \\ E & $-0.33
+0.19E-0.015E^2$ & $0.78+0.37E-0.03E^2$ & 1 \\ F & $-0.39
+0.19E-0.02E^2$ & $1.66+0.37E-0.035E^2$ & 1 \\ G & $ 0.5$ &
$1.53-0.32E-0.1E^2$ & 1 \\ H & $0.28 +0.29E -0.025E^2$ & $1.205$ & 1
\\ J & $0.47 $ & $1.0 -0.45 E$ & 1 \\ K & $0.29 +0.29E -0.03E^2$ &
$0.77-0.14E+0.013E^2$ & 1 \\
\end{tabular}
\end{ruledtabular}
\end{center}
\caption{\label{table_scans}Parameterization of energy-momentum scans
performed in Ref.~\onlinecite{Coldea3}: the momentum transfers ${\bf
k}=(h, k, l)$ are parameterized in terms of the energy transfer $E$
(in meV). $\hat{\bf k}_a $ is a measure of the polarization
factor. Given that the weak interlayer coupling is neglected, $h$ is
not needed for the purpose of our calculation.}
\label{table}
\end{table}

\begin{figure}[ht]
\includegraphics[width=8.6cm]{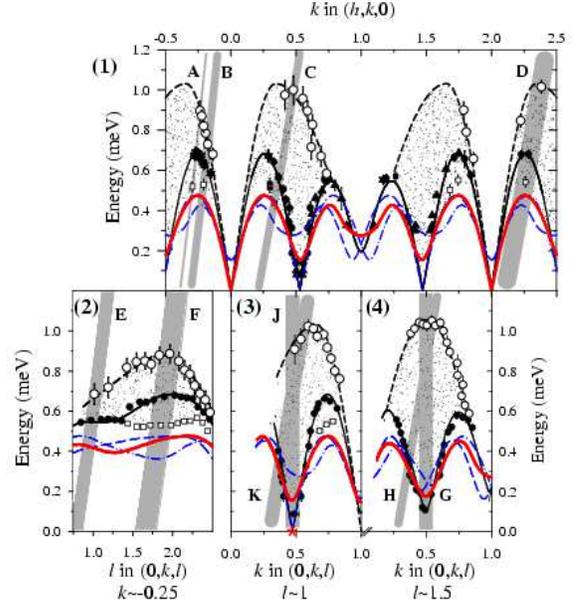}
\caption{(Color Online) The dispersion relation of magnetic
excitations. The shaded regions labeled with capital letters A through
K indicate scan directions (the line thickness indicates the wave
vector averaging). The filled symbols are the main peaks in the line
shape as determined experimentally in the ordered phase (from
Ref.~\onlinecite{Coldea3}). The dotted area indicates the extent of
the scattering continuum. The open circles and squares are
respectively the upper and lower boundary of the scattering continuum
as determined experimentally. The upper thick dashed line is a guide
to the eye. The thin solid line is the experimental fit to the
principal mode using effective parameters ($\tilde{J}=0.61$~meV,
$\tilde{J^{\prime}}=0.107$~meV). The thick solid, dashed and dash
dotted lines are respectively the $1/S$ results for the principal
($\omega^{0}_{\bf k}$) and secondary ($\omega^+_{\bf k}$,
$\omega^{-}_{\bf k}$) modes determined from Eq.~\ref{Dyson}. }
\label{EXPCURVES}
\end{figure}

\subsection{Excitation line shapes}
\begin{figure*}[ht]
\begin{picture}(17, 12)(0.2, 0.2)
\put(0, 6){\includegraphics[width=8.6cm]{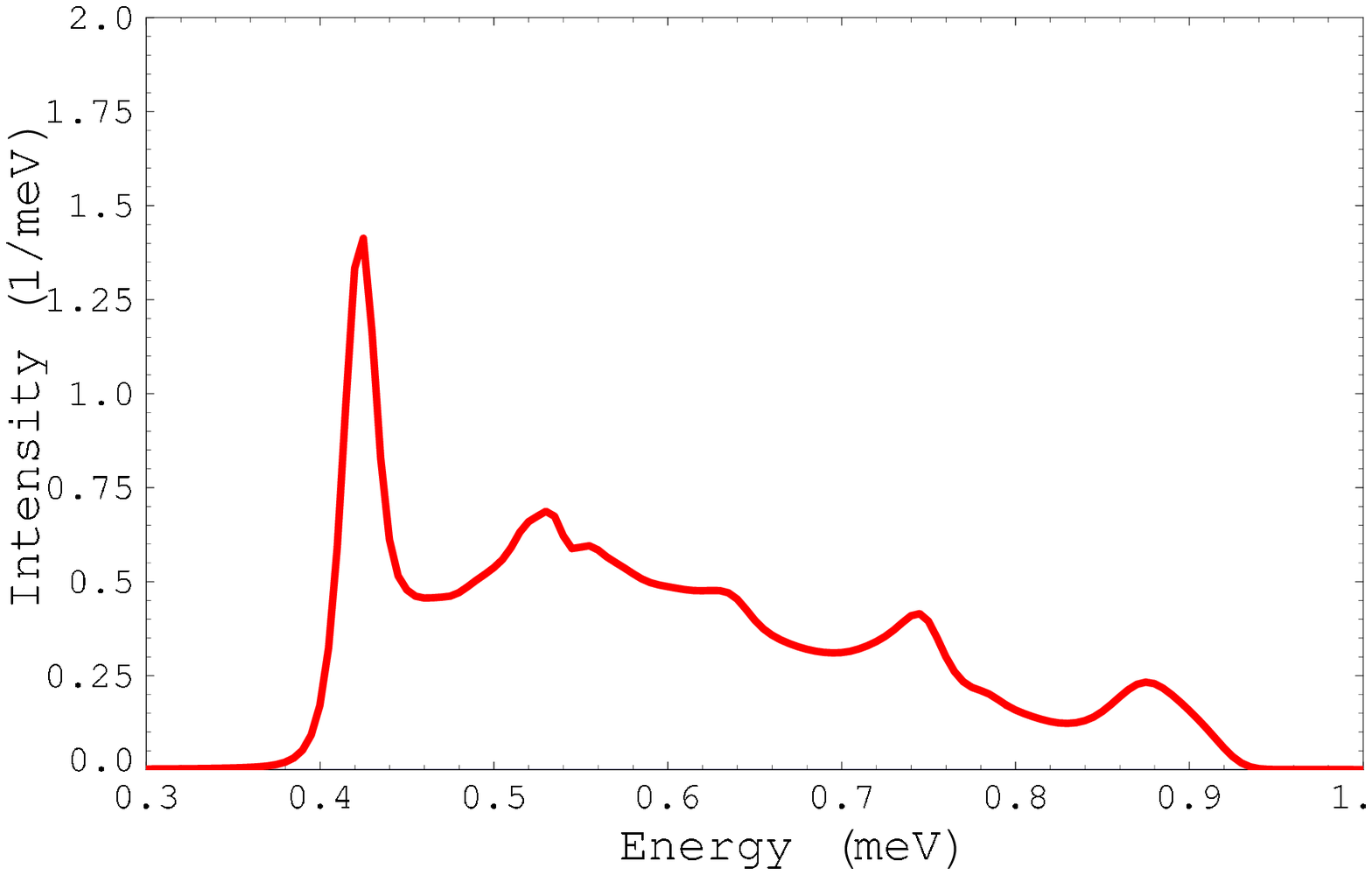}} \put(2.6,
7.7){\includegraphics[width=5.6cm]{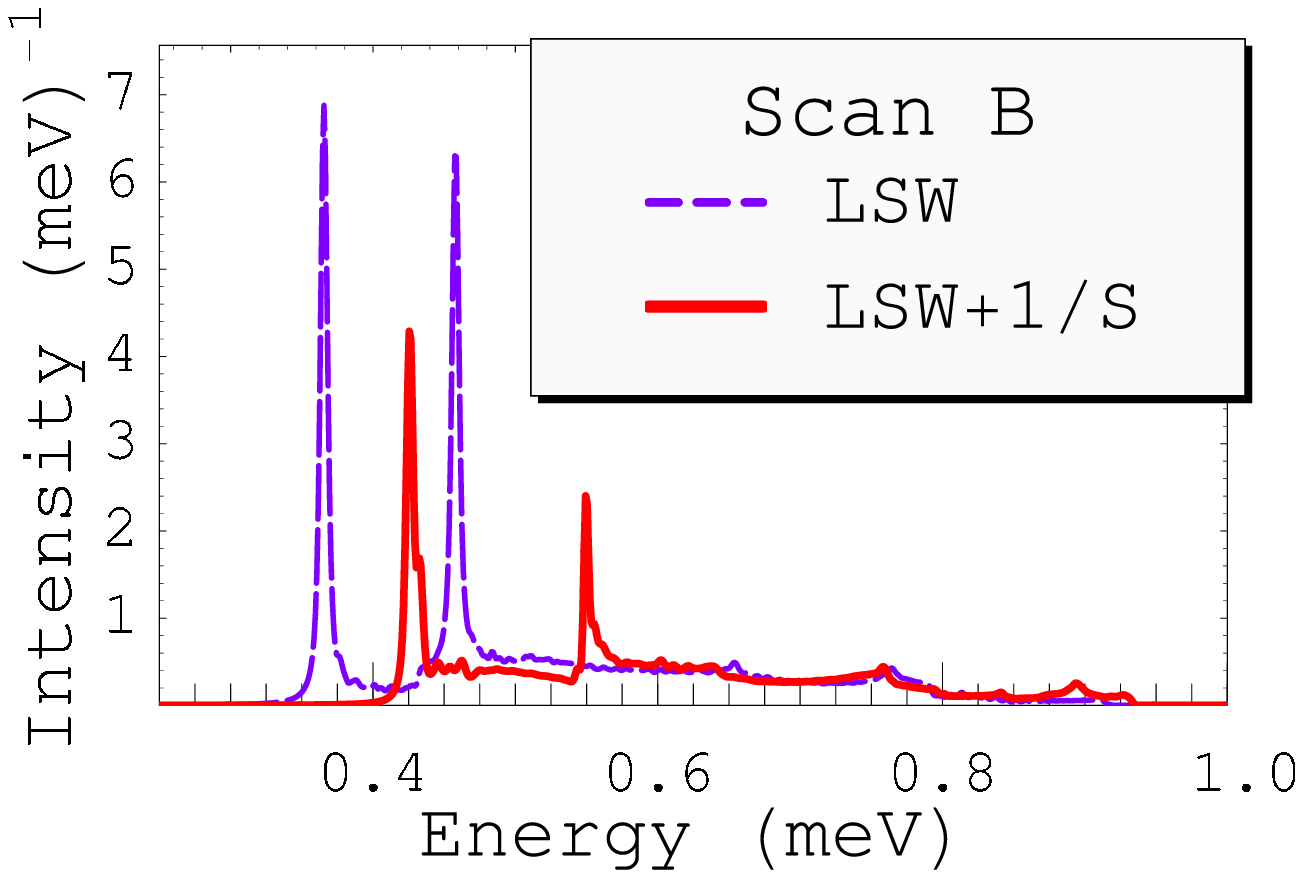}}
\put(8.6, 6){\includegraphics[width=8.6cm]{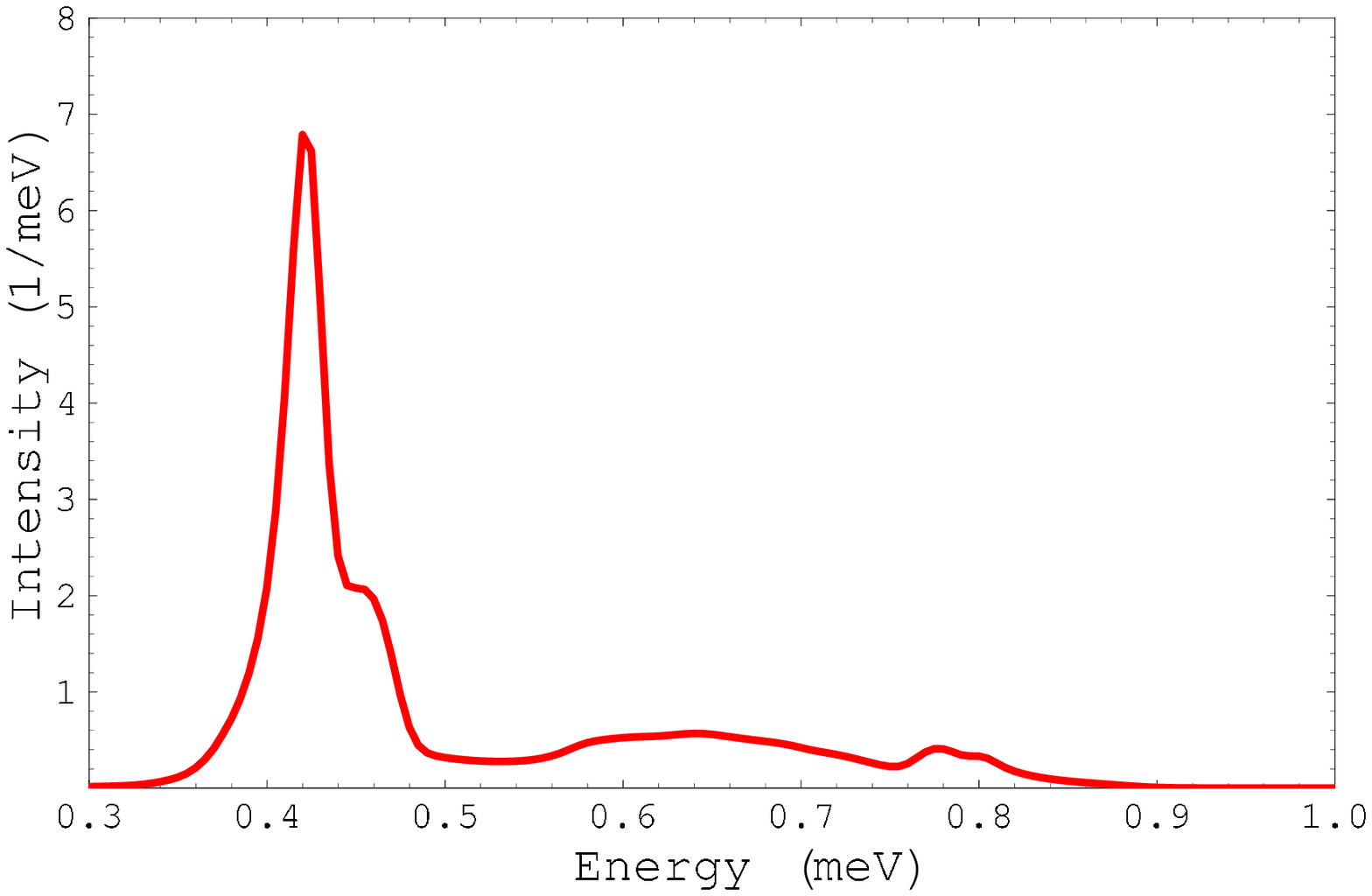}} \put(10.8,
7.4){\includegraphics[width=6.0cm]{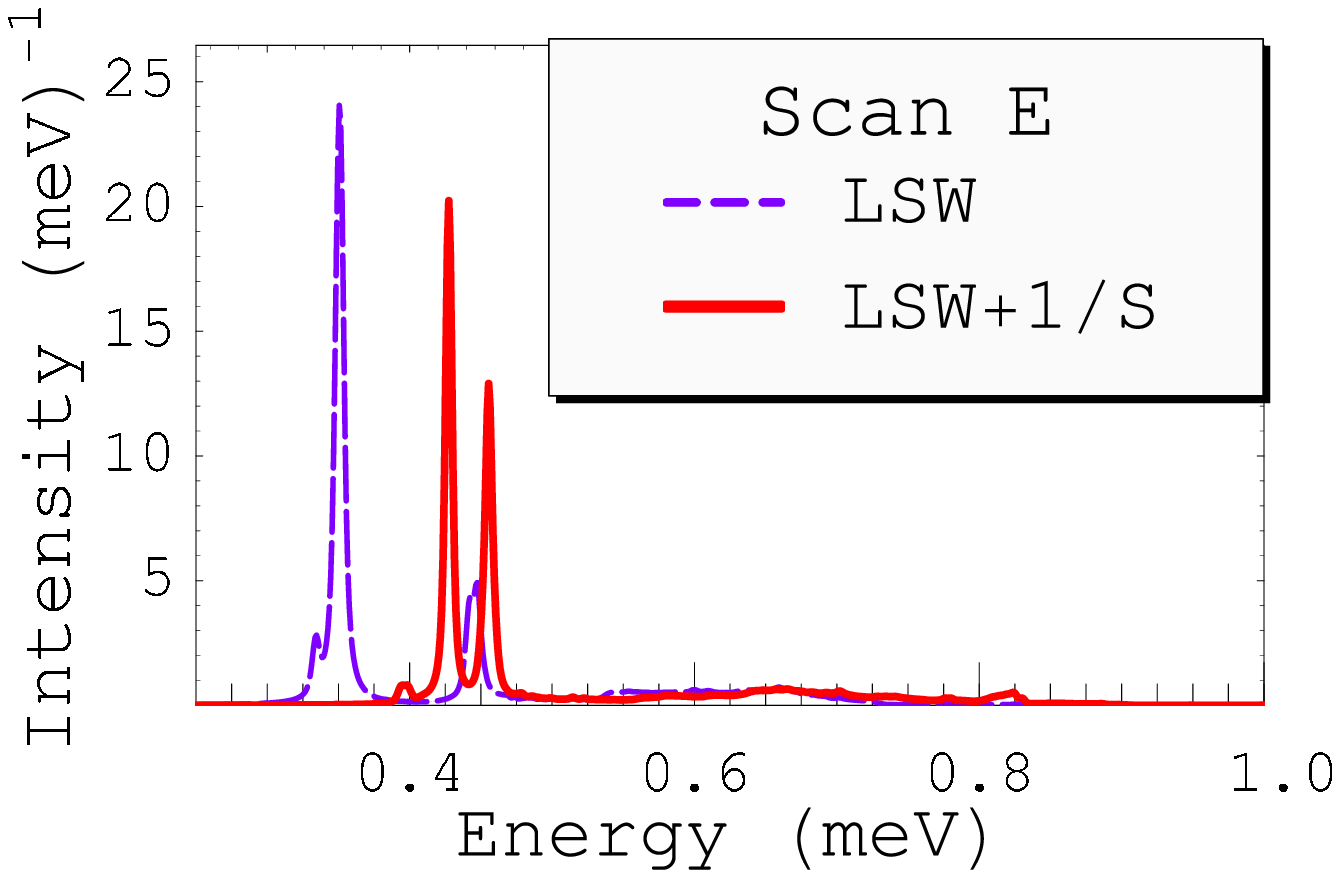}}
\put(0, 0){\includegraphics[width=8.6cm]{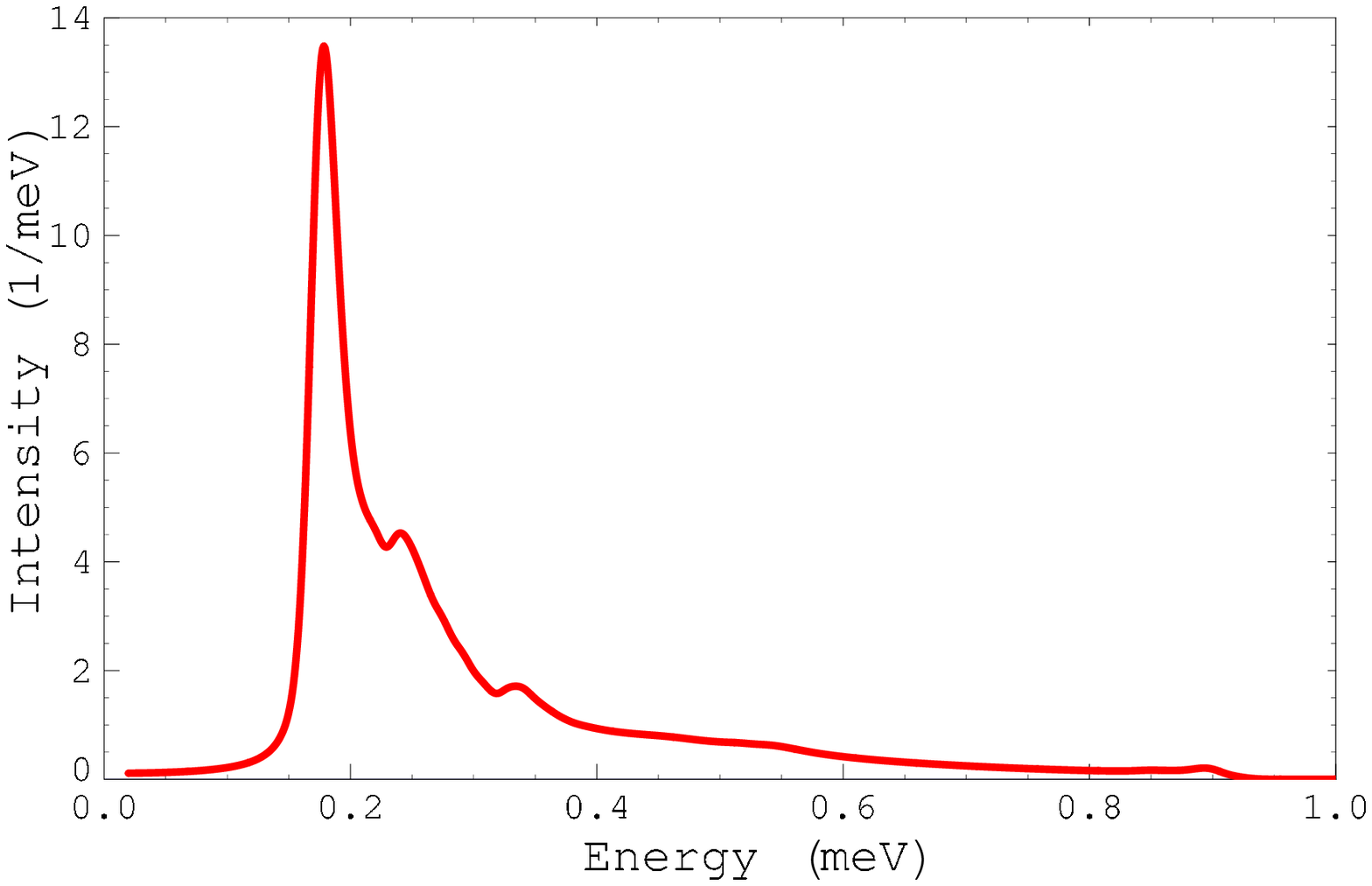}} \put(2.3,
1.6){\includegraphics[width=6.0cm]{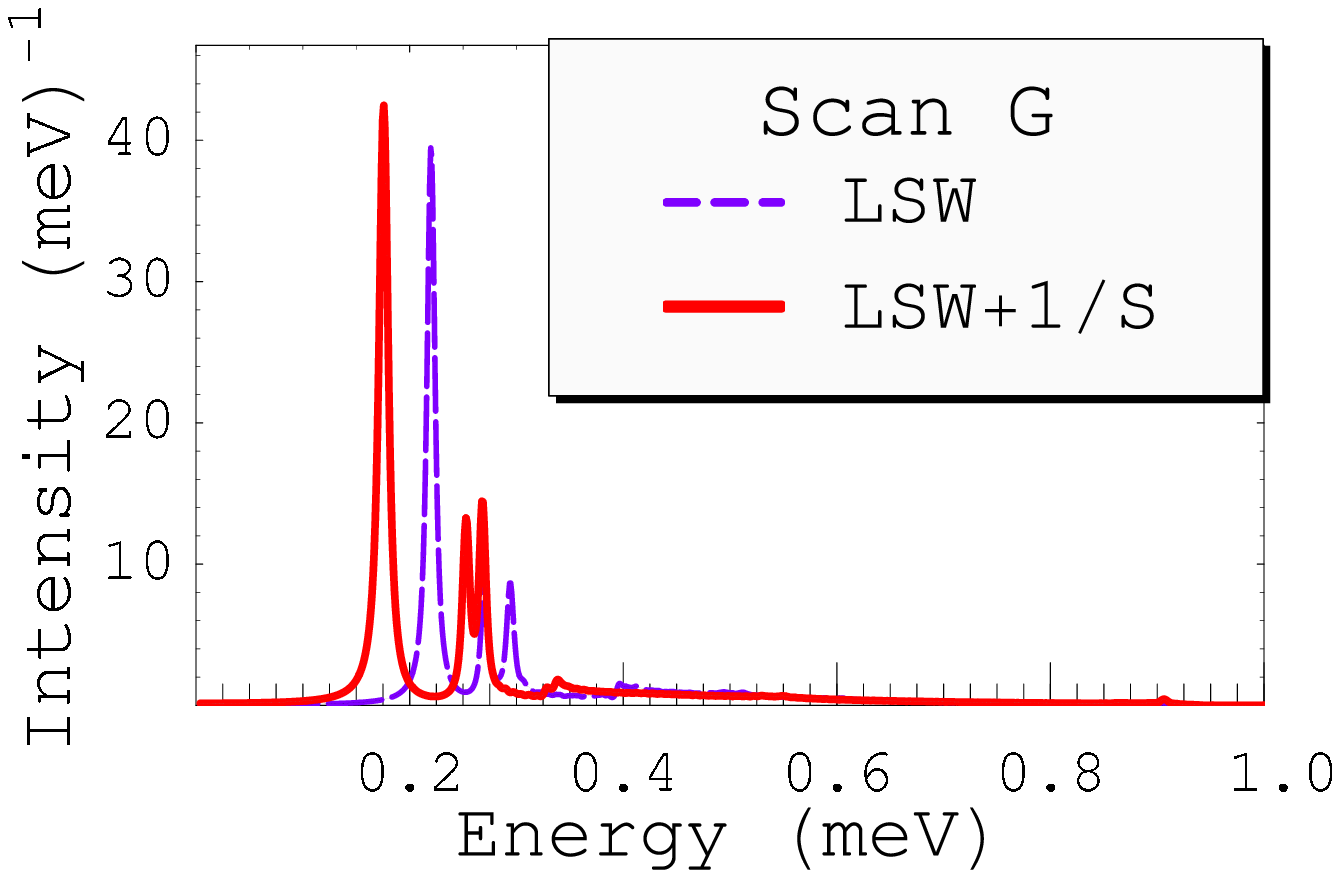}}
\put(8.6, 0){\includegraphics[width=8.6cm]{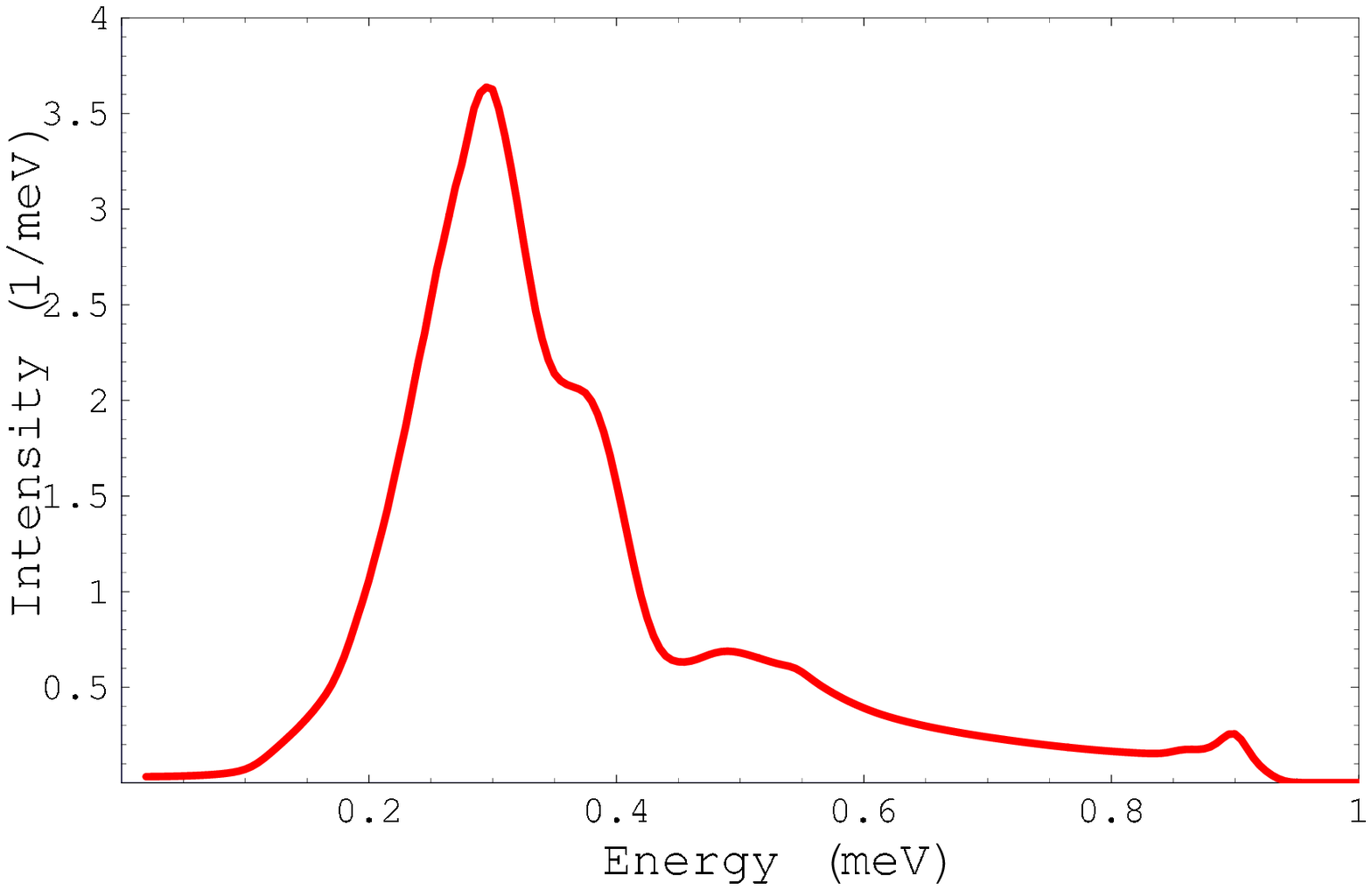}} \put(12.4,
1.8){\includegraphics[width=4.5cm, height=3.3cm]{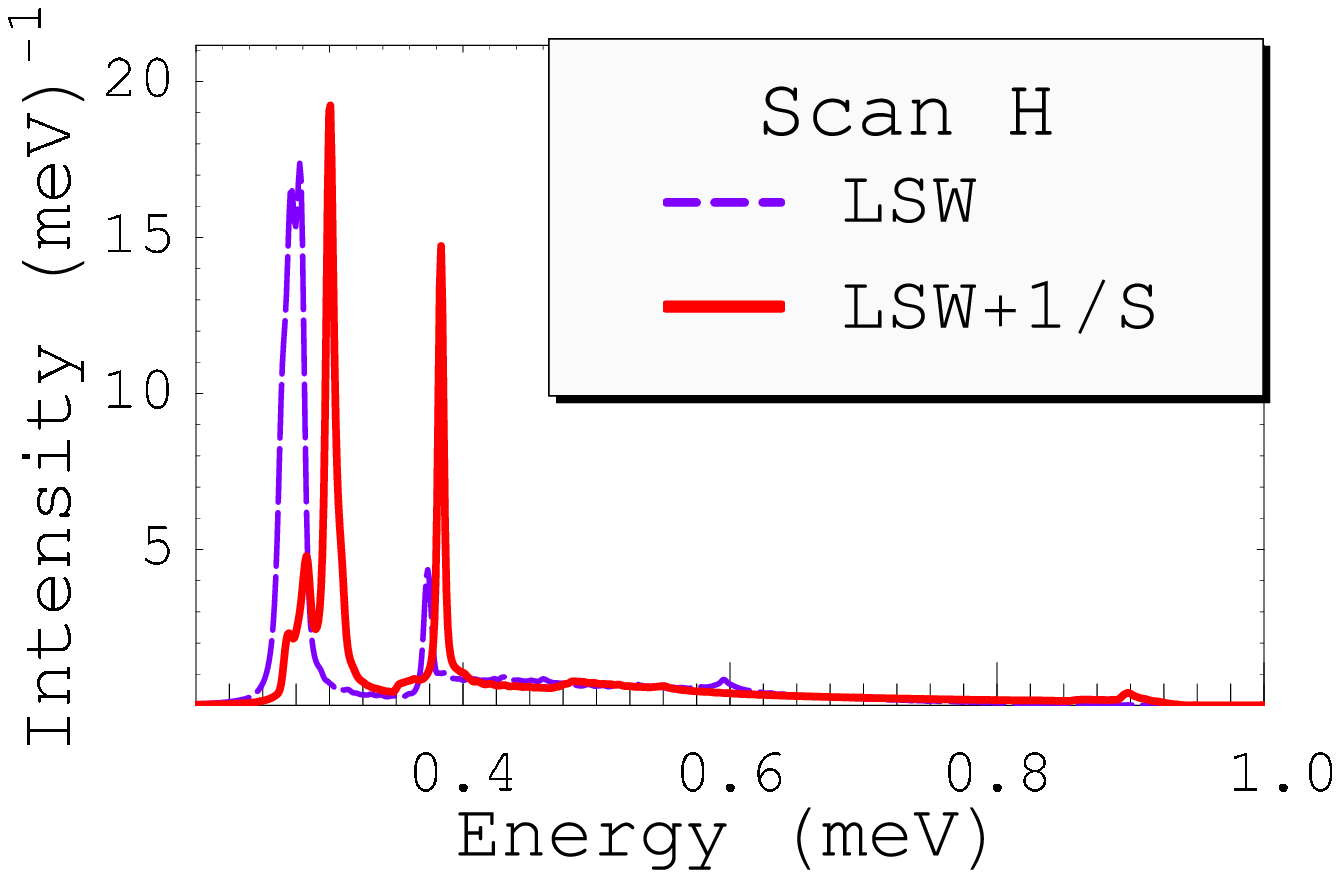}}
\put(1.5, 5){\makebox(0, 0){\textbf{ (3)}}} \put(10.5, 5){\makebox(0,
0){\textbf{ (4)}}} \put(1.5, 11){\makebox(0, 0){\textbf{ (1)}}}
\put(10.5, 11){\makebox(0, 0){\textbf{ (2)}}}
\end{picture}
\caption{ (Color Online) Scattering cross section. The numbered panels
(1-4) correspond to energy scans B, E, G and H respectively. The data
has been convolved with the energy and spatial resolution. $\Delta E
=0.016$~meV for all plots and $\Delta k=\{ 0.035, 0.039, 0.085, 0.056
\}$ for plots (1-4). The insets show the results of linear spin wave
(LSW) theory and the $1/S$ expansion (LSW+1/S) for $\Delta k=0, \Delta
E=0.002$ meV.}
\label{PLOTSCAN2}
\end{figure*}

\begin{figure}[h2t]
\begin{picture}(8.6, 3.6)(0.2, 0.2)
\put(0.1, 0){\includegraphics[width=4.3cm,height=3.5cm]{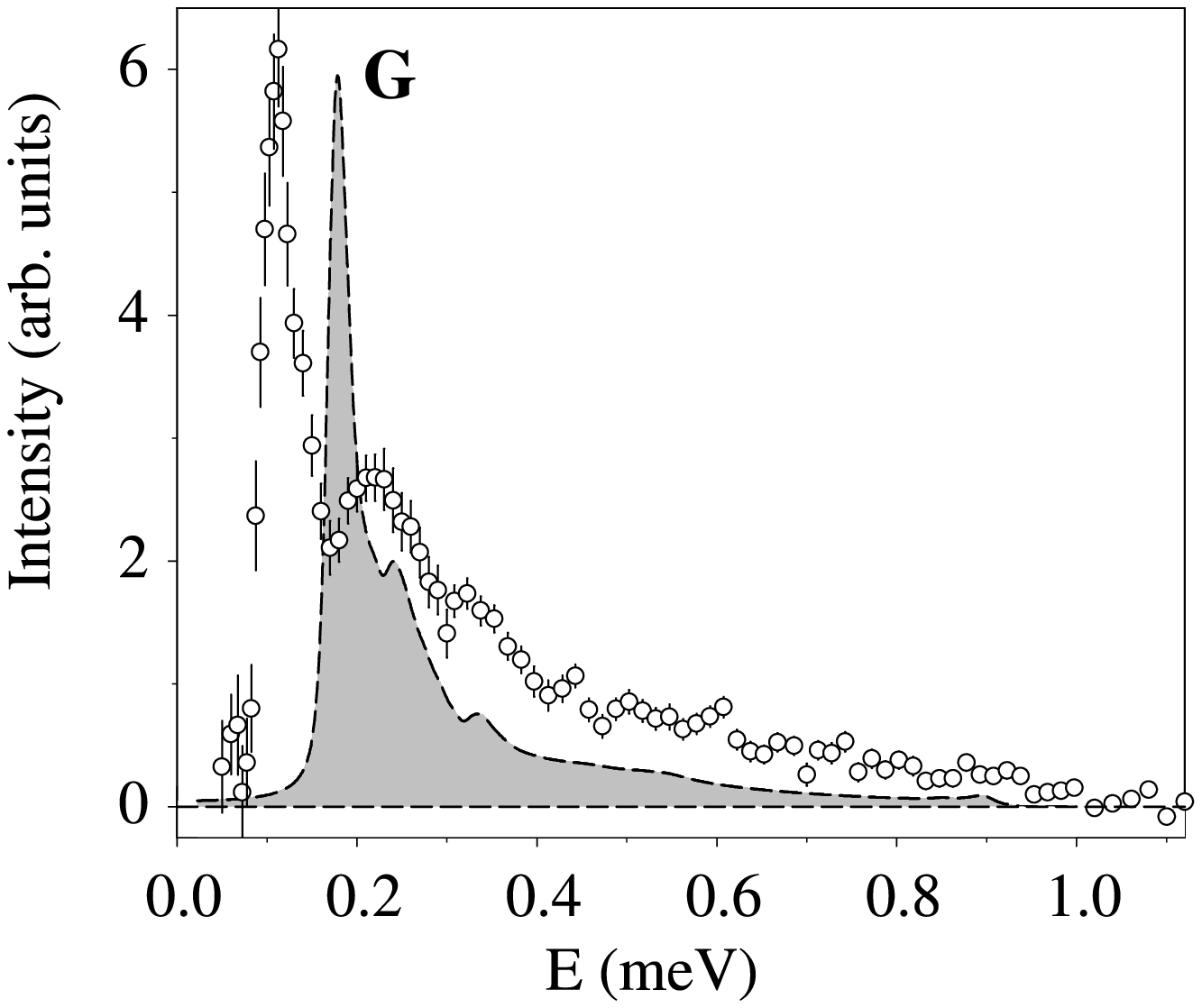}}
\put(4.4, 0){\includegraphics[width=4.3cm,height=3.5cm]{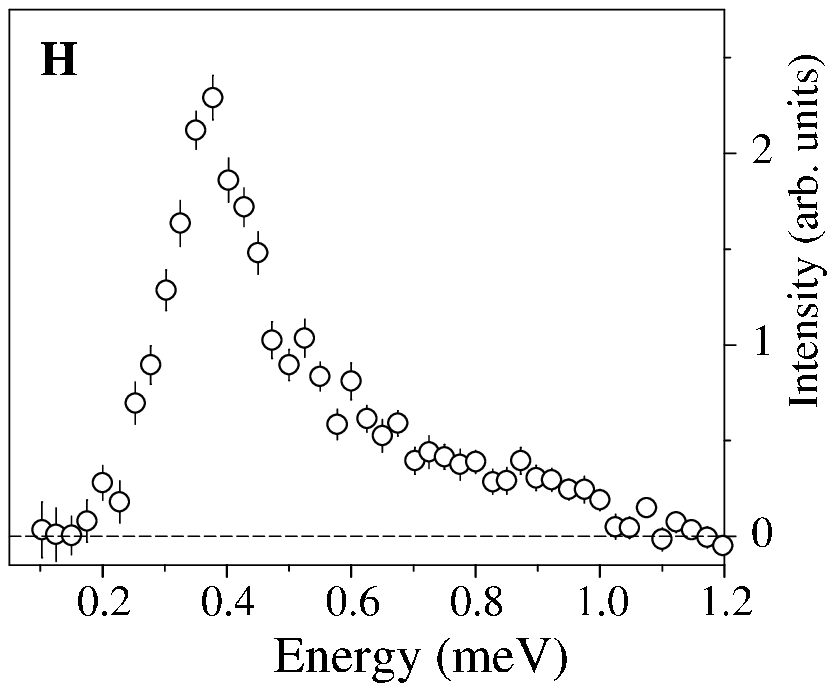}}
\put(2.2, 2.2){\makebox(0, 0){\textbf{ (1)}}} \put(6.5,
2.2){\makebox(0, 0){\textbf{ (2)}}}
\end{picture}
\caption{ Observed neutron scattering lineshape in scan G (1) and H
  (2) (data from Ref.~\onlinecite{Coldea3}) in the ordered phase
  ($T<0.1 K$). In scan G, the shaded area represents the                     
the 1/S calculation (Eq. (39)) convolved with the experimental                         
resolution.}
\label{EXPPLOTSCAN}
\end{figure}

\begin{figure}[h2t]
\begin{picture}(8, 6)(0.2, 0.2)
\put(0, 3){\includegraphics[width=4.2cm]{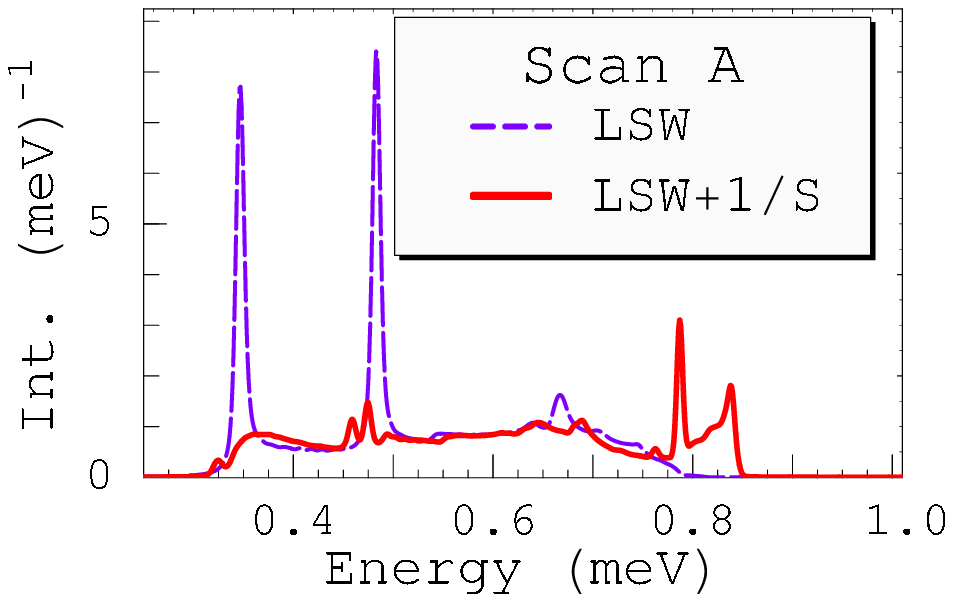}} \put(4,
3){\includegraphics[width=4.2cm]{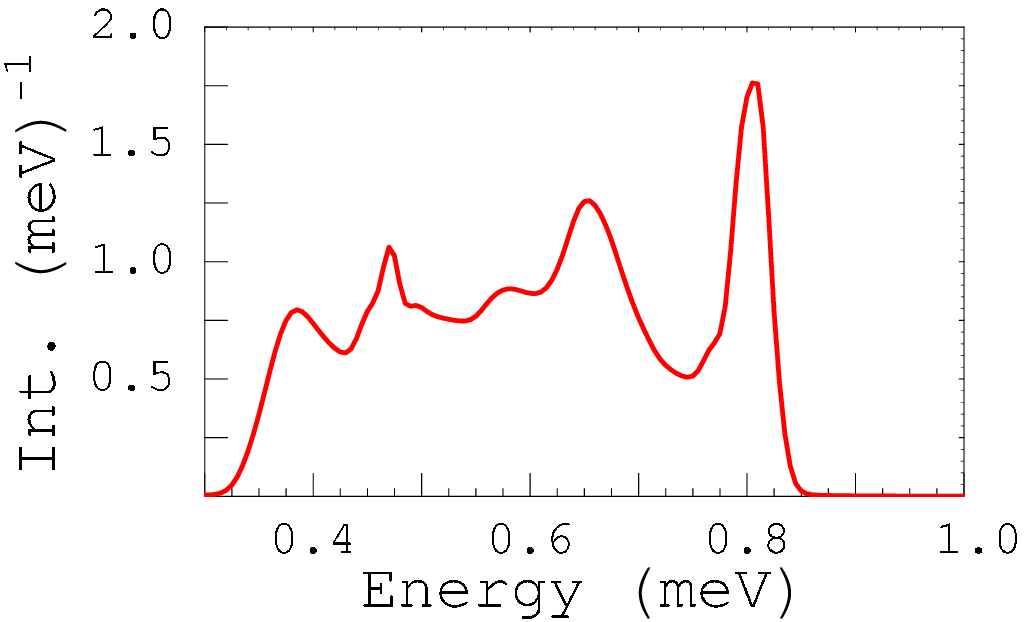}} \put(0,
0){\includegraphics[width=4.2cm]{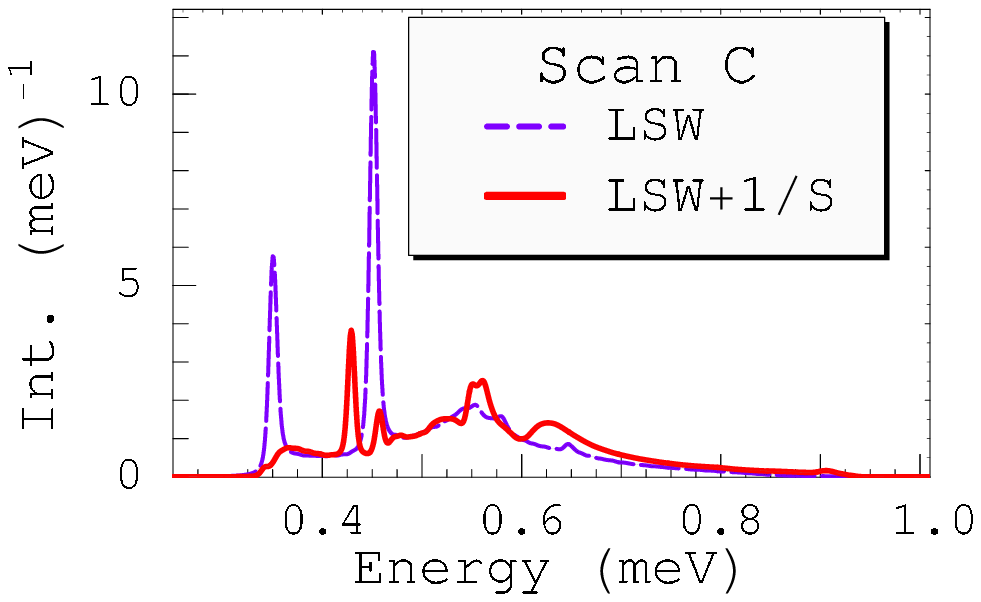}} \put(4,
0){\includegraphics[width=4.2cm]{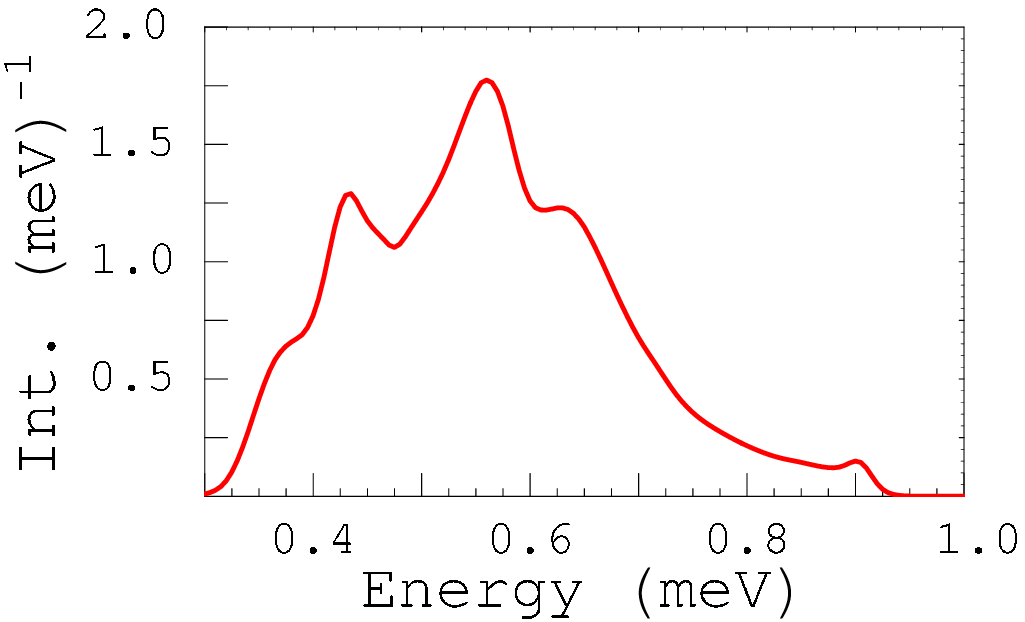}} \put(1.2,
2.2){\makebox(0, 0){\textbf{ (3)}}} \put(5.5, 2.2){\makebox(0,
0){\textbf{ (4)}}} \put(1.2, 5.2){\makebox(0, 0){\textbf{ (1)}}}
\put(5.5, 5.2){\makebox(0, 0){\textbf{ (2)}}}
\end{picture}
\caption{ (Color Online) Calculated scattering cross sections in
linear spin wave theory (LSW) and $1/S$ expansion (LSW+$1/S$) for
scans A and C. Panels (2) and (4) show the LSW+$1/S$ results with
instrumental resolutions of $\Delta E=0.016$~meV and $\Delta k=0.02$
(for scan A) and $\Delta k=0.04$ (for scan C) taken into account. See
Fig.~5 from Ref.~\onlinecite{Coldea3} to compare to experimental
data.}
\label{PLOTSCAN1}
\end{figure}

In order to exhibit the properties of the dynamical structure factor
in greater detail we have generated a series of scans in ${\bf
k}-\omega$ space. The inelastic neutron scattering measurements on
Cs$_2$CuCl$_4$ were not performed at constant momentum transfer but
followed various trajectories in energy-wave vector space. We have
generated our theoretical scans using the known parameterizations of
the scans A to J of Ref.~\onlinecite{Coldea3} in ${\bf k}-\omega$
space, which we summarize in Table.~\ref{table}. We refer the reader
to Ref.~\onlinecite{Coldea3} for further details. The various scans
are shown in Fig. \ref{EXPCURVES}. Also shown are the regions in which
significant magnetic scattering is observed experimentally and the
location of the main peaks. For comparison we plot the principal and
secondary spin wave dispersions obtained from the $1/S$ expansion. As
we have already emphasized, the $1/S$ expansion underestimates the
quantum renormalization of the exchange constants and as a result the
agreement of the calculated spin wave dispersions with the main peaks
observed experimentally is poor.

The experimental energy and momentum resolutions have been accounted
for to make contact with experiment. We find that the effects of the
finite energy resolution of $\Delta E=0.016$~meV are generally
outweighed by the effects of the finite momentum resolution. This is a
consequence of the large modulation of the spin wave dispersion along
the chain direction, i.e $(0k0)$, (whose slopes can reach
$\frac{\Delta E}{\Delta k} \sim 1.6$~meV), which causes an
amplification of the effects of the momentum resolution. Given that
the spin waves are nearly dispersionless along the $(00l)$ direction,
we have only taken into account the spatial resolution along the chain
direction.

To illustrate this point, let us consider the results for scans B, E,
G and H shown in Fig.~\ref{PLOTSCAN2}. The insets of panel (4) show
the results of both linear spin wave theory and the $1/S$ expansion
for a hypothetical energy resolution of $\Delta E=0.002$~meV which has
been introduced to make the various delta function peaks visible (the
momentum resolution is set to zero $\Delta k=0$). First we consider
the results for scan H (Panel 4 of Fig.~\ref{PLOTSCAN2}). Linear spin
wave theory predicts peaks at approximately $0.27$~meV and $0.37$~meV
corresponding to the degenerate spin wave modes $\omega^+, \omega^0$
and to $\omega^-$ respectively. The $1/S$ correction yields a slight
upward shift in the energy of these peaks. In both linear spin-wave
and $1/S$ calculations, the two magnon scattering continuum is found
to carry nearly a quarter of the integrated spectral weight. Taking
into account the finite momentum resolution (the width at half maximum
is $\Delta k=0.057$) we find that the sharp peaks get broadened very
significantly as is shown in panel (4). The dynamical structure factor
now exhibits an extended continuum in which the single-particle
excitation can no longer be resolved and merges smoothly with the two
magnon continuum. This result is qualitatively similar to the
experimental observations shown for comparison in panel (2) of
Fig. \ref{EXPPLOTSCAN}.

Next we turn to scan G (Panel 3 of Fig.~\ref{PLOTSCAN2}), which probes
the vicinity of the wave vector $(0, 0.5, 1.5)$. Experimentally a
resolution-limited peak is observed at an energy of $0.107(10)$~meV in
this region of intense scattering, see panel (1) of
Fig.\ref{EXPPLOTSCAN}. However, about two thirds of the spectral
weight is associated with a scattering continuum at higher
energies. Both linear spin wave theory and the $1/S$ expansion predict
sharp peaks in this region of the Brillouin zone. The $1/S$ expansion
gives a spin wave peak at $\omega^0=0.18$~meV carrying nearly half of
the spectral weight and two further peaks at energies around $0.25$
meV corresponding to the two secondary spin wave modes. The two magnon
scattering continuum extends up to $0.9$~meV and carries nearly a
quarter of the spectral weight. We emphasize that, in contrast to
$\omega^\pm$, the principal mode $\omega^0$ is close to a saddle point
and therefore is nearly dispersionless. In Panel 3 the finite energy
and momentum resolutions are taken into account. We see that the
almost dispersionless main mode remains sharp but the secondary modes
can no longer be resolved and are found to merge with the two
magnon continuum. Irrespective of the discrepancies between the
results of the $1/S$ expansions and the experimental data, our
calculation suggests that the lower boundary of 
of the measured scattering continuum in scan G could be due to
unresolved transverse magnons.  Such a scenario had been previously 
considered and ruled out on the basis of the smallness of the ratio 
$I_{\rm sec}/I_{\rm pri}$ of spectral weights of the secondary modes
to the principal mode predicted by linear spin wave theory.~\cite{Coldea3}
However, the results of the $1/S$ expansion show that spin wave
interactions lead to an enhancement of this ratio for the G scan. 

Next, we examine scan E (Panel 2), which probes wave vectors near
${\bf k}=(0, -0.25, 1)$. Linear spin wave theory predicts coherent
peaks at $\omega^0=0.35$~meV for the principal mode and at
$\omega^{-}=0.44$ meV and $\omega^{+}=0.33$~meV for the secondary
modes (see the inset in Panel 2). The two magnon scattering continuum
is relatively weak and carries only about 23 \% of the total spectral
weight. In the framework of the $1/S$ expansion the principal mode is
pushed upwards in energy to $\omega^0=0.42$~meV and occurs very close
to the secondary mode $\omega^{-}=0.45$~meV. The other secondary mode
$\omega^{+}$ is shifted very significantly to $0.39$~meV, but carries
only a minute fraction of the spectral weight. The two magnon
continuum is also shifted upwards in energy and carries approximately
a quarter of the total spectral weight. Once again the spin wave
dispersion is close to a saddle point and as a result the effects of
the finite momentum resolution are small. The main feature in the
structure factor is a broad peak formed by the two unresolved
$\omega^-$ and $\omega^{0}$ modes. This is quite similar to what is
observed experimentally (Fig.~5(E) of Ref.~\onlinecite{Coldea3}). It
is then tempting to speculate that the experimentally observed single
peak is a result of the accidental near degeneracy of the $\omega^{-}$
and $\omega^{0}$ modes in the vicinity of ${\bf k}=(0, -0.25,
1)$. This would explain both the absence of the $\omega^{-}$ peak in
the experimental data and the anomalously large intensity of the
observed peak.

In Panel 1 of Fig.~\ref{PLOTSCAN2} we plot the dynamical structure
factor for scan B near the wave vector $(2, -0.25, 0)$. Here the
polarization factor ($\hat{\bf k}_a $) in (\ref{crosssection}) leads
to a strong suppression of the out-of-plane fluctuations and the
scattering is almost entirely due to the in-plane $\omega^{\pm}$ spin
wave modes. The magnon interactions renormalize $\omega^{+}$ upwards
in energy to approximately $0.42$~meV, whereas the $\omega^{-}$ mode
disappears in the two magnon scattering continuum. A careful analysis
shows that the narrow peak at $0.55$~meV is not due to a
single-particle excitation but is a feature in the two magnon
scattering continuum.

The dominant contribution to the dynamical structure factor in scan A
in the vicinity of the wave vector $(1.5, -0.3, 0)$ comes from
in-plane fluctuations because the polarization factor $\hat{\bf k}_a $
suppresses out-of plane fluctuations. As can be seen in
Fig.~\ref{PLOTSCAN1}, the magnon interactions lead to a spectral
weight transfer to higher energies. The peaks near $0.8$~meV and
$0.85$~meV can be traced back to single-particle poles in the Green's
function. These poles are unphysical and are a result of the
uncontrolled nature of the $1/S$ expansion for small values of $S$. It
is easily seen from the Dyson equation (\ref{Green1}) that a large
self-energy at a given wave vector can lead to ``extra'' poles in the
Green's function at high energies above the two magnon continuum. The
inclusion of higher order terms in the $1/S$ expansion would provide
decay mechanisms at all energies and lead to a broadening of these
high-energy peaks in the dynamical structure factor.

Last but not least let us consider the vicinity of $(0.8, 0.4, 0)$
(scan C). As is shown in Fig.~\ref{PLOTSCAN1} the principal spin wave
mode $\omega^{0}$ is renormalized down to a slightly lower energy of
approximately $0.42$~meV. The $\omega^{+}$ mode, which occurs at
$0.35$~meV in linear spin wave theory, disappears entirely in the two
magnon-continuum. The feature near $0.60$~meV can again be understood
in terms of an enhancement of the two magnon density of
states. Comparing with the neutron scattering data (Fig.~5(C) of
Ref.~\onlinecite{Coldea3}), the structure factor shows features quite
similar to the experimentally observed continuum. However, the
scattering continuum occurs at energies nearly $0.10$~meV lower than
what is observed experimentally.


\section{Conclusions}
\label{Conclusions}

In this work we have used nonlinear spin wave theory to determine the
dynamical structure factor  in the ordered phase of the spin-1/2
helimagnet Cs$_2$CuCl$_4$.
We have taken into account the first subleading contribution in a
$1/S$ expansion, which incorporates interactions between magnons and
generates magnon decay processes as well as multi magnon scattering
continua. Both effects are particularly pronounced in Cs$_2$CuCl$_4$
due to the non-collinear spin ordering, the low spin value and
geometrical frustration. 

We found that the results of nonlinear spin wave theory explain on a
qualitative level many of the features observed in neutron scattering
experiments. We find a strong scattering continuum in the dynamical 
structure factor similar to the experimental observations. 
Our calculations suggest the possibility that some of the spectral
weight at the low-energy boundary of the experimentally observed
scattering continuum in scan G could be due to single particle
excitations that are unresolved. 

In the vicinity of saddle points
of the spin wave dispersion relation the single-particle excitations
are only weakly affected by the instrumental resolution and hence
exhibit sharper peaks in the dynamical structure factor.

In spite of the qualitative agreement of the theory with experiments,
crucial discrepancies remain. First and foremost, nonlinear spin wave
theory fails to account for the large ``quantum renormalization'' of
the main exchange parameter. This indicates that (to order ${\cal
O}(S^0)$) the $1/S$ expansion still underestimates the effects of
quantum fluctuations. 
Furthermore, there are significant quantitative differences between
our calculations and the experimentally observed structure factor. 
One may speculate that a better agreement with experiment could be
achieved by taking higher-order terms in the $1/S$ expansion into account. 

The main lesson to be learned from our calculations is that
Cs$_2$CuCl$_4$ falls somewhere in between the two theoretical
scenarios that have been proposed previously. Our analysis shows that
the physics of order plays an essential part in understanding the
dynamic response Cs$_2$CuCl$_4$ at low temperatures: a large fraction
of the spectral weight is carried by spin wave modes, which occur over
a large range of frequencies. This is a strong indication that a
putative spin-liquid ground state is plainly not a good starting point
for the description of the ordered phase of Cs$_2$CuCl$_4$. On the
other hand we have seen that (in low orders in $1/S$) nonlinear spin
wave theory significantly underestimates the effects of quantum
fluctuations and hence expansions around the ordered state also fail
to account for the experimental observations.

Nonlinear spin wave theory can also be applied to investigate the
effects of magnetic fields. It is known that in the presence of a
field linear spin wave theory is generally a very poor approximation
as it excludes all-important magnon decay
processes.\cite{Zhitomirsky01} A self-consistent study of magnetic
field effects in Cs$_2$CuCl$_4$ is currently under way.\cite{EJV}
During completion of this work, we became aware of a parallel effort
which reaches similar conclusions.\cite{OtherWork}

\acknowledgments{ The work was supported by the EPSRC under Grant
GR/R83712/01. We are grateful to John Chalker and Alan Tennant for
valuable discussions. Particular thanks are due to Radu Coldea for
numerous helpful discussions and suggestions as well as providing us
with figures \ref{EXPCURVES} and \ref{EXPPLOTSCAN}.}



\end{document}